\documentclass[11pt,a4paper]{article} 
\usepackage{jheppub,hyperref,mdwlist}
\usepackage{jheppub,hyperref,mdwlist}
\usepackage{amsmath}
\usepackage{graphicx}
\usepackage{subfigure}
\usepackage{amssymb}
\usepackage{xcolor}
\usepackage{multirow}
\usepackage{cancel}
\usepackage{color}
\usepackage{mathtools}
\usepackage{listings}
\usepackage{xcolor}
\usepackage{float}
\usepackage{slashed}
\usepackage{bm}
\usepackage{amsmath}
\usepackage{wrapfig}

\newcommand{\be}{\begin{equation}}
\newcommand{\ee}{\end{equation}}
\newcommand{\beq}{\begin{equation}}
\newcommand{\eeq}{\end{equation}}
\newcommand{\bea}{\begin{eqnarray}}
\newcommand{\eea}{\end{eqnarray}}
\definecolor{airforceblue}{rgb}{0.36, 0.54, 0.66}
\definecolor{steelblue}{rgb}{0.27, 0.51, 0.71}
\definecolor{amber}{rgb}{1.0, 0.49, 0.0}

\title{\boldmath Interplay between dark matter and leptogenesis in a common framework}

\author{XinXin Qi,}
\author{Hao Sun}
\affiliation{Institute of Theoretical Physics, School of Physics, Dalian University of Technology, No.2 Linggong Road, Dalian, Liaoning, 116024, P.R.China }
\emailAdd{qxx@mail.dlut.edu.cn}
\emailAdd{haosun@dlut.edu.cn}

\abstract{
We consider the interplay between dark matter and leptogenesis in a common framework, 
where three right-handed neutrinos, one fermionic dark matter and two singlet scalars are introduced into the Standard Model. 
The mixing of the two singlet scalars not only determines the dark matter relic density but also connects right-handed neutrino with dark matter. 
We consider that the baryon asymmetry is generated via the resonant leptogenesis and the right-handed neutrino masses are at the TeV level. 
We consider a viable parameter space satisfying the relic density constraint, 
and the parameter space is more flexible in the case of a larger mixing angle. 
We found that the existence of dark matter in the model can not only dilute the baryon asymmetry 
but can also generate a larger baryon asymmetry due to the process of dark matter annihilation into a pair of right-handed neutrinos. 
Both the dilution effect and enhanced effect can occur so that influence the observed baryon asymmetry.
}

\begin{document}
\maketitle
\flushbottom

\section{Introduction}
\label{sec:intro}

The observed baryon asymmetry and dark matter relic density in the universe~\cite{Planck:2018vyg} 
are two long-standing problems that can provide us with new hints for new physics. 
Moreover, the baryon asymmetry can be obtained through the leptogenesis mechanism~\cite{Fukugita:1986hr,Luty:1992un,Plumacher:1996kc,Buchmuller:2004nz,Davidson:2008bu}, 
which is also related to the problem of tiny neutrino masses. 
As for the dark matter, the correct relic density can be produced conservatively by Freeze-out~\cite{Chiu:1966kg} or Freeze-in~\cite{Hall:2009bx} 
depending on the strength of the interactions associated with dark matter and the initial density of dark matter in the early universe. 
   
A common framework for unified dark matter as well as leptogenesis can be found in Ref.~\cite{Josse-Michaux:2011sjn}, 
which can explain quantitatively both the observed baryon asymmetry of our universe and dark matter relic density. 
However, this extension is difficult to be falsified. 
On the other hand, according to the fact of $\Omega h^2 \approx 5 \Omega_b h^2$~\cite{Planck:2018vyg}, 
where $\Omega h^2 \approx 0.12 $ is the observed dark matter relic density and $\Omega_b h^2$ is the baryon density, 
a shared mechanism to generate baryon asymmetry and dark matter simultaneously will be much more attractive for the quantitive relationship, 
and related discussions can be found in Refs.~\cite{Chu:2021qwk,Cui:2020dly,Cui:2016rqt,Dasgupta:2016odo,Cui:2012jh}. 
Especially, when comes to leptogenesis, the right-handed neutrinos may not only couple to the Standard Model (SM) 
but also a hidden sector where the dark matter resides, and a lepton asymmetry as well as dark matter production 
can be both generated by the out-of-equilibrium decay of the right-handed neutrinos~\cite{Falkowski:2011xh,Falkowski:2017uya}. 
However, such scenarios often indicate asymmetric dark matter (ADM)~\cite{Nussinov:1985xr,Kaplan:1991ah,Zurek:2013wia,Graesser:2011wi}, 
where relic dark matter is not determined by the annihilation cross section but asymmetry between particle-antiparticle number densities of dark matter. 
   
From the point of the standard leptogenesis and WIMP (weakly interacting massive particles) DM scenario, it seems that there is little connection between leptogenesis and dark matter. On the one hand, a lepton asymmetry is generated by the decay of right-handed neutrino (RHN) out of equilibrium, then the lepton asymmetry is converted into baryon asymmetry by the sphaleron process. Such a process always happens at a high scale and demands the right neutrino mass larger than $10^9$ GeV  with Davidson-Ibarra bound \cite{Davidson:2002qv} while dark matter is still in thermal equilibrium. On the other hand, dark matter will make little difference in baryon asymmetry result in the case of Freeze-in due to the weak interaction. Alternatively, low-scale leptogenesis will be more attractive because high-scale leptogenesis is difficult to be tested at colliders. The degenerate right-handed neutrino mass can contribute to a resonant enhancement to the CP violation so that one can obtain the baryon asymmetry with right-handed mass at the TeV scale, which is the so-called resonant leptogenesis \cite{Pilaftsis:2003gt}. Other low-scale leptogenesis scenarios such as the Akhmedov-Rubakov-Smirnov (ARS) mechanism can be found in \cite{Hambye:2016sby,Alanne:2018brf,Liu:2020mxj}. Although we can decrease right-handed neutrino mass to a low scale, dark matter may still be hardly relevant to leptogenesis since the dark matter will freeze out earlier than leptogenesis in the case of dark matter mass much larger than right-handed neutrino mass. 
   
   In fact, the symmetric WIMP DM can be connected with baryon asymmetry via the so-called WIMPy miracle \cite{Cui:2011ab}, where the WIMP DM annihilation is directly responsible for baryogenesis. In the WIMPy baryogenesis, the Sakharov conditions are satisfied with: baryon number violation, CP violation and departure from thermal equilibrium, then a non-zero baryon number asymmetry can be generated from dark matter annihilation, and one can obtain the observed baryon asymmetry as well as dark matter relic density simultaneously. Furthermore, the WIMPy leptogenesis has also been discussed in \cite{Kumar:2013uca}, in which the lepton asymmetry arises from dark matter annihilation processes which violate CP and lepton number, and the lepton asymmetry is then converted into baryon asymmetry by electroweak sphalerons.
  
  In this paper, we will not focus on the new mechanism to generate the baryon asymmetry and dark matter relic density but  focus on the interplay between dark matter and baryon asymmetry. Therefore, we discuss them in a conservative choice that dark matter relic density is generated by the Freeze-out mechanism and baryon asymmetry is obtained via resonant leptogenesis, and the latter means right-handed neutrino masses are degenerate. Although baryon asymmetry is not generated by dark matter annihilation with the WIMPy miracle, the existence of dark matter can affect the baryon asymmetry result as long as dark matter production is related to right-handed neutrinos. We consider the case that right-handed neutrino masses are approximate to the dark matter mass. Such a region will be much more interesting since either dark matter or right-handed neutrinos are out-of-equilibrium and dark matter Freeze-out as well as leptogenesis can occur nearly at the same time. Generally speaking, new processes related to right-handed neutrinos will keep right-handed neutrino number density close to equilibrium so that dilute the baryon asymmetry. However, dark matter can also annihilate into a pair of right-handed neutrinos therefore baryon asymmetry will be strengthened. Notice that right-handed mass can be at the TeV scale in the case of successful resonant leptogenesis, and we demand the dark matter mass also at the TeV scale.
  
  We will not give a UV-completion model in this work, but a minimal scenario including right-handed neutrino and dark sector. For the dark sector, we introduce a singlet scalar $S$ and a fermion $\chi$, where $\chi$ carries $i$ charge and $S$ with $-1$ charge under discrete $Z_4$ symmetry. $S$ obtains vaccum expectation value and $\chi$ acquires mass after spontaneously breaking. For the visible sector, we introduce three right-handed neutrinos $N_j$ ($j=1,2,3$) as well as a singlet scalar $\phi$ to the Standard Model (SM). The singlet $\phi$ also obtains the non-zero vev and couples to the three right-handed neutrinos with $\phi N_i N_j$. As for the scalar $\phi$, we have the following comments. Firstly, the singlet-neutrino couplings break the accidental global lepton-number symmetry of the SM, which is similar to the right-handed neutrino Majorana masses in the seesaw Lagrangian. As mentioned in Ref.~\cite{Alanne:2018brf}, this indicates such Yukawa couplings are possible to present in any low energy effective theory that contains the singlet scalar $\phi$ and a set of sterile neutrinos. Then, we assume the singlet $\phi$ is odd under a $Z_2$ symmetry of the scalar potential, such $Z_2$ symmetry can be the resident symmetry of  a new gauge symmetry such as $U(1)_{B-L}$ after spontaneously breaking and the $Z_2$ symmetry can limit the couplings in the scalar potential. On the other hand, the $Z_2$ symmetry in the scalar potential may be at most an approximate symmetry that is explicitly broken by the singlet-neutrino couplings due to the small couplings. Last but not least, the introduction of the $Z_2$ symmetry can decrease the input parameters in the model and simplify our discussion. $\phi$ does not couple to $\chi$ directly because of the different discrete symmetry and we introduce $S$ to the model. As a result, the mixings of the  scalars can induce processes related to dark matter and right-handed neutrinos. We consider the decoupling limit that both mixing angles of the singlet scalars with SM Higgs can be negligible so that  contribution of the SM Higgs to the dark matter production can be ignored, and dark matter relic density is related to the mixing of the singlet scalars.
  
The paper is arranged as followed, we describe our framework in section \ref{sec:2}. In section \ref{sec:3}, we give the Blotzman equations related to baryon asymmetry and dark matter. In section \ref{sec:4}, we give the evolution of dark matter with different parameters. In section \ref{sec:5}, we scan the parameter space the satisfying dark matter relic density constraint as well as baryon asymmetry constraint and discuss the relationship between dark matter and baryon asymmetry. We give a summary in the last part.

\section{The model framework}
\label{sec:2}

In this part, we give the framework including dark matter and right-handed neutrino. 
For the dark sector, we introduce one singlet scalar $S$ and a fermion $\chi$, 
where $\chi$ as the dark matter carries $i$ charge under a discrete $Z_4$ symmetry and $S$ charge is $-1$. 
$S$ can obtain non-zero vev $v_s$ after spontaneously symmetry breaking (SSB) so that $\chi$ will acquire mass. 
We also introduce three right-handed neutrinos and another singlet scalar $\phi$ to the Standard Model. 
The additional Lagrangian is given as followed,
\begin{eqnarray} \label{eq1}
 \mathcal{L} &=& - y\bar{L}NH - \frac{1}{2}M_n \bar{N^c}N- \frac{1}{2}\lambda_{mn} \phi \bar{N^c}N -\frac{1}{2} \lambda_{sx} S \bar{\chi^c}\chi  -\mathcal{V}(S,H,\phi),
\end{eqnarray}
suppressing the generation indexes, where $L$ is the SM leptons and $H$ is the SM Higgs doublet. 
The term $\mathcal{V}(S,H,\phi)$ in Eq.~\ref{eq1} involves Higgs potential and other terms with the singlet scalar $S$ as well as $\phi$. 
We assume the singlet $\phi$ is $Z_2$ odd in the scalar potential and obtain the vaccum expectation value $v_b$ after SSB. 
Therefore, the term $\mathcal{V}(S,H,\phi)$ can be given by,
 \begin{eqnarray} \nonumber
  \mathcal{V}(S,H,\phi)&=&-\mu H^2+\lambda H^4-\mu_s S^2+\lambda_s S^4-\mu_p \phi^2 \\\nonumber
                       &+& \lambda_p \phi^4 + \lambda_{hs} H^2 S^2 +\lambda_{hp}H^2\phi^2 \\
                       &+& \lambda_{sp}S^2\phi^2 .
 \end{eqnarray}
Note that we have no terms such as $H^2S\phi$ because $S$ and $\phi$ carry different charges. 
In the unitary gauge, $H$, $S$ and $\phi$ can be given by,
   \begin{eqnarray}
   H= \left(
   \begin{array}{c}
   0\\
   \frac{v+h}{\sqrt{2}}
   \end{array}
   \right), ~~
   S=\frac{v_s+s}{\sqrt{2}}, ~~
   \phi =\frac{v_b+\tilde{\phi}}{\sqrt{2}} ,
   \end{eqnarray}
where $v=246$ GeV is the SM vaccum expectation value.
We consider the decoupling limit that $\lambda_{hp}, \lambda_{hs} \ll 1$ so that the mass matrix $M_h$ for scalars after SSB can be simplified by,
 \begin{eqnarray}
   M_h= \left(
   \begin{array}{ccc}
   2\lambda_s v_s^2 & \lambda_{sp}v_bv_s & 0\\
   \lambda_{sp}v_bv_s&2\lambda_p v_b^2 & 0\\
   0 &0 &2\lambda v^2 \\
   \end{array}
   \right).
   \end{eqnarray}
 The mass matrix $M_h$ can be diagonalized by the matrix $U$ with
 \begin{eqnarray}
 U= \left(
   \begin{array}{ccc}
  \cos\theta & \sin\theta & 0\\
  -\sin\theta & \cos\theta & 0\\
   0 &0 &1 \\
   \end{array}
   \right)
   \end{eqnarray}
where $\theta$ is the mixing angle of the two singlet scalars. Correspondingly, after rotation from the flavor eigenstate to the mass eigenstate, we have  three Higgses defined by,
   \begin{eqnarray}
  \left(
   \begin{array}{c}
  h_1\\
  h_2 \\
  h_0 \\
   \end{array}
   \right)=
   \left(
   \begin{array}{ccc}
  \cos\theta & \sin\theta & 0\\
  -\sin\theta & \cos\theta & 0\\
   0 &0 &1 \\
   \end{array}
   \right)
   \left(
   \begin{array}{c}
  s\\
  \tilde{\phi} \\
  h \\
   \end{array}
   \right),
   \end{eqnarray}
where $h_0$ is the SM Higgs, and $h_{1,2}$ are the new Higgs particles in the model. 
Particularly, $h_1$ and $h_2$ can mix with each other, and the mixing angle $\theta$ will also determine dark matter relic density. 
Furthermore, we choose Higgs masses as well as the mixing angle as the inputs so that parameters related to singlet scalars in the model can be expressed as follows,
\begin{eqnarray} \nonumber
&&\lambda_s = \frac{m_1^2\cos^2\theta}{2v_s^2}+\frac{m_2^2\sin^2\theta}{2v_s^2} \\ \nonumber
&&\lambda_p = \frac{m_1^2\sin^2\theta}{2v_b^2}+\frac{m_2^2\cos^2\theta}{2v_b^2} \\
&&\lambda_{sp} =  \frac{(m_2^2-m_1^2)\sin 2\theta}{2v_bv_s}
\end{eqnarray}
 where $m_{1,2}$ are the $h_{1,2}$ mass. To obtain a stable potential, we have
 \begin{eqnarray}
 4\lambda_{s}\lambda_p >\lambda_{sp}^2, \lambda_{s,p}>0,
 \end{eqnarray}
known as co-positivity constraints~\cite{Kannike:2012pe}, and the perturbativity constraints are given by $\lambda_{sp,s,p}<4\pi$.
 
We consider a conservative choice of the decoupling limit that the mixings of singlets with SM doublets are negligible. 
Therefore, the contribution of SM particles to the dark matter production such as $ff \to \chi\chi$ are highly suppressed where $f$ represents the SM fermions, 
and the left relevant processes with dark matter are the right-handed neutrinos as well as the $h_{1,2}$. 
It is worth stressing that the most stringent bounds on the mixing angle $\alpha$ of the SM Higgs 
arise from the $W$ boson mass correction~\cite{Lopez-Val:2014jva}, and the current constraint is given by $|sin\alpha| \lesssim  0.24$ at $95\% $ C.L. according to \cite{Papaefstathiou:2022oyi}.

In this work, we assume the vaccum expectation value $v_b$ as a free parameter and both $v_b$ and $v_s$ are beyond EWPT (electroweak phase transition) scale. For simplicity, we fix $v_b=2$ TeV as a constant in the following discussion. 
After SSB, dark matter mass $m_{\chi}$ can be given by $m_{\chi}=\lambda_{sx}v_s/2\sqrt{2}$, 
the term $(M_n + \lambda_{mn}v_b/\sqrt{2})/2$ gives the Majarona mass matrix of the right-handed neutrino 
with $M_n$ being the bare mass term of right-handed neutrinos, 
and the dynamical origin of $M_n$ is left unspecified for our work. 
The matrix $\lambda_{mn}$ characterizes the strength of the Yukawa coupling of $\phi NN$ 
while $\lambda_{sx}$ describes the strength of singlet-DM Yukawa interaction. 
  
  Notice that the existence of the singlet scalar $\phi$ can also induce a successful low-scale leptogenesis by the scalar-singlet-mediated one-loop diagrams due to the Yukawa coupling with right-handed neutrinos \cite{Alanne:2018brf} only if $M_n$ and $\lambda_{mn}$ can not be diagonalized simultaneously, and we ignore the effect of $\phi$ on leptogenesis in our work. 
  
  In this work, we consider the interplay between dark matter and leptogenesis in a minimal framework, we assume dark matter relic density is generated via the Freeze-out mechanism, and dark matter production processes related to SM particles are highly suppressed under the decoupling limit, while almost contribution to DM relic density comes from the new Higgses as well as right-handed neutrinos. On the other hand, we consider the observed baryon asymmetry is generated by resonant leptogenesis, and annihilation of right-handed neutrinos to dark matter may keep the neutrino number close to thermal equilibrium which dilutes the baryon asymmetry. Inversely, more right-handed neutrinos can be generated by the inverse process so that the baryon asymmetry will be strengthened. In both cases, the final baryon asymmetry can be influenced by the introduction of dark matter.

\section{Boltzmann equations}
\label{sec:3}

In this section, we discuss the Boltzmann equations of $N$ and dark matter. 
In our work, we consider the decoupling limit that the mixings of both singlets with SM doublets are negligible. 
Under such consideration, dark matter production is mainly related to right-handed neutrinos as well as $h_{1,2}$, 
while the contribution from SM particles is highly suppressed. 
Extra contribution to the leptogenesis can arise from the channels $NN \to \chi \chi$, which is also related to dark matter. Notice that DM can annihilate into pairs of scalars that are heavier than the DM particles, which is similar with the so-called  forbidden DM scenario \cite{DAgnolo:2015ujb}. On the other hand, we assume $m_N <m_{1,2}$ and ignore the contribution of  $NN \to h_{1,2}h_{1,2}$ to the BAU.
 
The Boltzmann equations for the $N$ abundance $Y_N$, dark matter abundance $Y_{X}$ and the total $(B-L)$ asymmetry $Y_{B-L}$ are given by \cite{Liu:2021akf},
\begin{eqnarray} \label{eq4} \nonumber
&&  \frac{s_NH_N}{z^4}\frac{dY_N}{dz}=-(\frac{Y_N}{Y_{Neq}}-1)(\gamma_D +2\gamma_{hs}+4\gamma_{ht})-(\frac{Y_N^2}{Y^2_{Neq}}-\frac{Y_X^2}{Y_{Xeq}^2})2\gamma_{N\chi}  \\ \nonumber
&&  \frac{s_NH_N}{z^4}\frac{dY_{B-L}}{dz}=-(\frac{Y_{B-L}}{2Y_{Leq}}+\epsilon_{CP}(\frac{Y_N}{Y_{Neq}}-1))\gamma_D -\frac{Y_{B-L}}{Y_{Leq}}(2(\gamma_N +\gamma_{Nt}+\gamma_{ht})+\frac{Y_N}{Y_{Neq}}\gamma_{hs}) \\
&&  \frac{s_NH_N}{z^4}\frac{dY_X}{dz}=-(\frac{Y_X^2}{Y_{Xeq}^2}-\frac{Y_N^2}{Y^2_{Neq}})2\gamma_{N\chi}-(\frac{Y_X^2}{Y_{Xeq}^2}-1)2\gamma_{\chi h} 
\end{eqnarray}
where $\epsilon_{CP}$ is the CP asymmetry parameter. The CP asymmetry $\epsilon_i$ can be given by \cite{Iso:2010mv}:
 \begin{align}
 \label{epsilon}
\epsilon_i= &\frac{\sum_j\Gamma_{N_i\to\ell_jH}-\Gamma_{N_i\to\bar\ell_jH^*}}{\sum_j\Gamma_{N_i\to\ell_jH}+\Gamma_{N_i\to\bar\ell_jH^*}}\\
=&-\sum_{j\neq i}\frac{m_{N_i}\Gamma_{N_j}}{m_{N_j}^2}\left(\frac{V_{ij}}{2}+S_{ij}\right)\frac{{\rm Im}(yy^\dagger)^2_{ij}}{(yy^\dagger)_{ii}(yy^\dagger)_{jj}},
\end{align}
where
\begin{align}
V_{ij}=&~2\frac{m_{N_j}^2}{m_{N_i}^2}\left[\left(1+\frac{m_{N_j}^2}{m_{N_i}^2}\right)\ln\left(1+\frac{m_{N_j}^2}{m_{N_i}^2}\right)-1\right],\\
S_{ij}=&~\frac{m_{N_j}^2(m_{N_j}^2-m_{N_i}^2)}{(m_{N_j}^2-m_{N_i}^2)^2+m_{N_i}^2\Gamma_{N_j}^2},
\end{align}
are respectively the vertex correction and RHN self-energy correction to the decay process with $m_{N_i}$ the $N_i$ mass and $\Gamma_{N_i}$ the $N_i$ decay width.  On the other hand, the tiny left-handed neutrino masses $m_{\nu}$ are generated via the Type-I seesaw mechanism with \cite{Liu:2021akf}:
 \begin{eqnarray}
  m_{\nu} \sim \frac{yv^2}{m_N} \sim 0.06 \mathrm{eV} \times (\frac{y}{10^{-6}})^2(\frac{1\mathrm{TeV}}{m_N}),
 \end{eqnarray}
 For the $\mathcal{O}(\mathrm{TeV})$ leptogenesis, we have $y \sim 10^{-6}$, and this makes the  contribution of terms $\gamma_{hs,ht,N,Nt}$ rather small since they are proportional to $y^2$ or $y^4$. 

The BAU obtains contributions from three right-handed neutrinos when we assume the right-handed neutrinos are degenerate. If the decay widths of the three right-handed neutrinos are comparable, then the generated BAU should be three times  that mere one right-handed neutrino decay. On the other hand, if the decay widths have a hierarchy, the CP asymmetries will also do so and the generated BAU can be dominated by one of the right-handed neutrinos \cite{Iso:2010mv}, so that a one-flavor discussion will be sufficient, and we consider such a scenario in this work
 and we have $\epsilon_{CP}=\epsilon_1$. For simplicity, we assume that dark matter is also determined by mere one right-handed neutrino, and we will have a similar conclusion when we consider three right-handed neutrinos. We denote $N$ as $N_1$ and $\lambda_{mn}$ as the Yukawa coupling of $\phi N_1N_1$ for simplicity.
 $z$ is defined by $z=m_N/T$ with $T$ being the temperature. $H_N$ and $s_N$ correspond to the Hubble rate and entropy density respectively. 
$Y_{Neq}$ as well as $Y_{Xeq}$ correspond to the abundance of right-handed neutrino and dark matter at thermal equilibrium respectively, 
which can be given by, 
  \begin{eqnarray} \nonumber
  &&Y_{Neq}(z)=\frac{45z^2}{2\pi^4 g_*}K_2(z), \\
  &&Y_{Xeq}(z)=\frac{45z^2m_{\chi^2}}{m_N^22\pi^4 g_*}K_2(z\frac{m_{\chi}}{m_N})
 \end{eqnarray}
and $Y_{Leq}$ is the lepton abundance at thermal equilibrium with
 \begin{eqnarray}
 Y_{Leq}=\frac{6}{s_N}\frac{m_N^3\zeta(3)}{4\pi^2}
 \end{eqnarray}
 where $g_{*}=106.75$ is the effective degree of freedom, $\zeta(x)$ is the Riemann zeta function and $K_2(x)$ is the Bessel function.
 
The term $\gamma_D$ is the reaction rate for $N \to HL$, 
$\gamma_{hs}$ and $\gamma_{ht}$ are the reaction rate of s-channel $NL\to q t$ 
and t-channel $Nt \to Lq$ mediated by Higgs, where $t$ is the top quark. 
The terms $\gamma_N$ and $\gamma_{Nt}$ in the $(B-L)$ asymmetry equations 
are the s-channel and t-channel contributions of $LH \to\bar{L}H$, which can wash out the baryon asymmetry. 
We follow the results of \cite{Iso:2010mv}, 
and the analytic expressions for these terms can be found in \cite{Iso:2010mv}. 
We have new terms involving dark matter as well as RHN. 
The term $\gamma_{N\chi}$ represents the reaction rate of $NN \to \chi\chi$, which is defined by,
   \begin{eqnarray}\label{eq12}
   \gamma_{N\chi}=\frac{m_N}{8\pi^4 z}\int_{max\{4m_N^2,4m_{\chi}^2\}}^{\infty}dx\hat{\sigma_{N\chi}}(x)x^{\frac{3}{2}}K_1(\frac{z}{m_N}\sqrt{x})~~
   \end{eqnarray}
where $K_1(x)$ is the modified Bessel function, 
$x$ is the squared center-of-mass energy 
and $\hat{\sigma_{N\chi}}(x)$ is the reduced cross section of $NN \to \chi \chi$, which is given by,
    \begin{eqnarray}\label{eq:8} \nonumber
    \hat{\sigma_{N\chi}}(x) &=& \frac{1}{4}\theta(\sqrt{x}-2m_N)\lambda^2(1,\frac{m_N^2}{x},\frac{m_N^2}{x})\frac{\cos^2\theta\lambda_{sx}(m_1^2-m_2^2)^2(x-4m_N^2)\sin^2\theta y_{mn}^2}{8\pi(x-m_1^2)^2(x-m_2^2)^2} \\
    &\times&\sqrt{\frac{(x-4m_{\chi}^2)(x-4m_N^2)}{x^2}}
    \end{eqnarray}
where $\lambda(a,b,c)=\sqrt{(a-b-c)^2-4bc}$, $\theta(\sqrt{x}-2m_N)$ is the theta function.
Note that the contribution of right-handed neutrinos to dark matter can be ignored in the limit of $\sin 2\theta\to 0$ 
so that dark matter relic density is  determined by the new Higgses, and dark matter will make no difference in the baryon asymmetry.
  
 \begin{figure}[h]
 \centering
 \includegraphics[width=12cm,height=3cm]{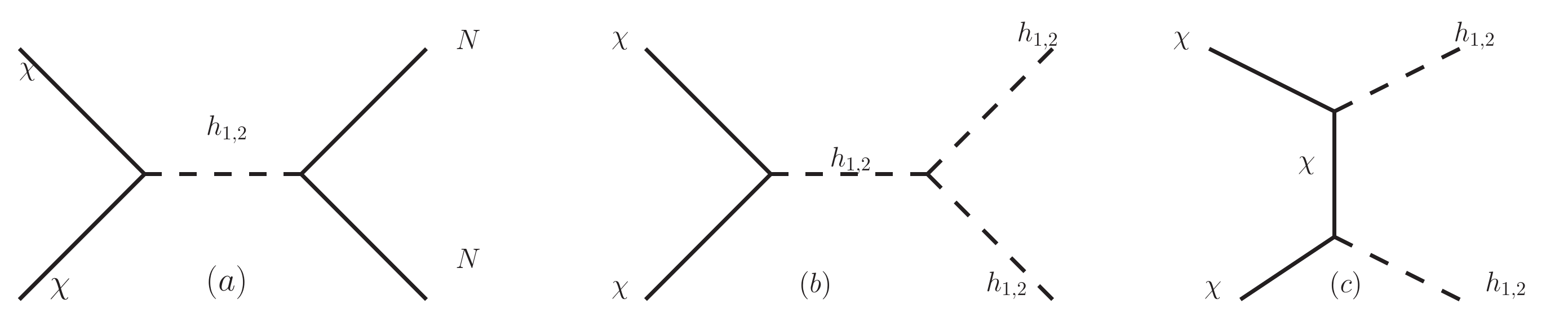}
 \caption{Processes related with dark matter relic density.}\label{Fig1}
\end{figure}  
The term $\gamma_{\chi h}$ represents the sum of the reaction rate for $\chi\chi\to h_1h_1, h_2h_2$ and $h_1h_2$. 
Relevant processes are given in Fig.~\ref{Fig1}, we have the following results of the reduced cross section, 
    \begin{eqnarray} \nonumber
    &&\hat{\sigma}_{\chi\chi\to h_1h_1}(x)=\frac{1}{4}\theta(\sqrt{x}-2m_{\chi})\lambda^2(1, \frac{m_{\chi}^2}{x}, \frac{m_{\chi}^2}{x})\sigma_{11} \\\nonumber
    &&\hat{\sigma}_{\chi\chi\to h_1h_2}(x)=\frac{1}{4}\theta(\sqrt{x}-2m_{\chi})\lambda^2(1, \frac{m_{\chi}^2}{x}, \frac{m_{\chi}^2}{x})\sigma_{12} \\
    &&\hat{\sigma}_{\chi\chi\to h_2h_2}(x)=\frac{1}{4}\theta(\sqrt{x}-2m_{\chi})\lambda^2(1, \frac{m_{\chi}^2}{x}, \frac{m_{\chi}^2}{x})\sigma_{22} ~~~~~~~~
    \end{eqnarray} 
where $\sigma_{ij}$ corresponds to the cross section of $\chi\chi\to h_ih_j$ with $i,j=1,2$. 
The cross sections related to the processes of Fig.~\ref{Fig1} are calculated with Calchep~\cite{Belyaev:2012qa}, 
and the expressions involving only s-channels are given as follows, while the complete expressions can be found in App.~\ref{App}.
\begin{eqnarray} \nonumber
   \sigma_{11}&=&\frac{1}{128m_{\chi}^2\pi(x-m_1^2)^2(x-m_2^2)^2v_b^2}\sqrt{\frac{(x-4m_1^2)(x-4m_{\chi}^2)}{s^2}}\sin^2\theta  \\ \nonumber
   & \times & ( 3\sqrt{2}m_2^2m_{\chi}^2(x-m_2^2)\sin^5\theta + 2\cos^5\theta \lambda_{sx}(m_2^2-m_1^2)m_{\chi}(x+2m_1^2)v_b \\ \nonumber
   &+& 2\cos^3\theta\lambda_{sx}(m_2^2-m_1^2)m_{\chi}(x+2m_1^2)\sin^2\theta v_b +\sqrt{2}\cos^2\theta \sin^3\theta(3m_{\chi}^2(m_2^2x+m_1^2(x-2m_2^2)) \\ 
   &+&2\lambda_{sx}^2(m_1^2-m_2^2)(m_1^2-3m_2^2+2x)v_b^2))^2
\end{eqnarray}
\begin{eqnarray} \nonumber
  \sigma_{22} &=& \frac{1}{128m_{\chi}^2\pi(x-m_1^2)^2(x-m_2^2)^2v_b^2}\sqrt{\frac{(x-4m_2^2)(x-4m_{\chi}^2)}{x^2}}\cos^2\theta \\ \nonumber
  &\times& (3\sqrt{2}m_1^2m_{\chi}^2(x-m_1^2)\cos^5\theta -2\cos^2\theta \lambda_{sx}(m_1^2-m_2^2)m_{\chi}(x+2m_2^2)v_b\sin^3\theta \\  \nonumber
  &-&2\lambda_{sx}(m_1^2-m_2^2)m_{\chi}(x+2m_2^2)\sin^5\theta v_b +\sqrt{2}\cos^3\theta \sin^2\theta(3m_{\chi}^2(m_2^2x+m_1^2(x-2m_2^2)) \\
  &+& 2\lambda_{sx}^2(m_2^2-m_1^2)(m_2^2-3m_1^2+2x)v_b^2))^2
\end{eqnarray}
and
\begin{eqnarray} \nonumber
  \sigma_{12}&=&\frac{\cos^2\theta \sin^2\theta}{64m_{\chi}^2\pi(x-m_1^2)^2(x-m_2^2)^2v_b^4}\sqrt{\frac{(x-4m_{\chi}^2)(m_1^4+(x-m_2^2)^2-2m_1^2(m_2^2+x))}{x^3}} \\\nonumber
  &\times& (3\sqrt{2}m_{\chi}^2(\cos^2\theta m_1^2+ m_2^2\cos^2\theta) (\cos^2\theta(m_1^2-x)+(m_2^2-x)\sin^2\theta) \\\nonumber
  &-& 2\cos\theta\lambda_{sx}(m_1^2-m_2^2)(\cos^4\theta(m_1^2-x) +2\cos^2\theta(m_2^2-m_1^2)\sin^2\theta \\
  &+& (x-m_2^2)\sin^4\theta)v_b^2)^2.
\end{eqnarray}
For the reaction rate of $\chi\chi\to h_1h_1,h_1h_2$ and $h_2h_2$, we have 
   \begin{eqnarray}    \label{eq12} 
   \gamma_{\chi h_1}&=&\frac{m_N}{8\pi^4 z}\int_{max\{4m_1^2,4m_{\chi}^2\}}^{\infty}dx\hat{\sigma}_{\chi\chi \to h_1h_1}(x)x^{\frac{3}{2}}K_1(\frac{z}{m_N}\sqrt{x}) \\ 
   \gamma_{\chi h_1h_2}&=&\frac{m_N}{8\pi^4 z}\int_{max\{(m_1+m_2)^2,4m_{\chi}^2\}}^{\infty}dx\hat{\sigma}_{\chi\chi \to h_1h_2}(x)x^{\frac{3}{2}}K_1(\frac{z}{m_N}\sqrt{x}) \\
   \gamma_{\chi h_2}&=&\frac{m_N}{8\pi^4 z}\int_{max\{4m_2^2,4m_{\chi}^2\}}^{\infty}dx\hat{\sigma}_{\chi\chi \to h_2 h_2}(x)x^{\frac{3}{2}}K_1(\sqrt{x}\frac{z}{m_N}).
   \end{eqnarray}
Note that we have ignored the contribution of $\chi\chi\to \nu\nu$ and $\chi\chi\to N\nu$ due to the tiny heavy-light neutrino mixing angle, where $\nu$ represents neutrino. 
On the other hand, the interactions involving dark matter as well as $h_{1,2}$ do not enter the baryon asymmetry equation at this order and therefore can not wash out the asymmetry.
    
We assume our universe started with the total $(B-L)$ charge zero, 
and the non-zero baryon asymmetry $Y_B$ can be dynamically generated 
above the sphaleron decoupling temperature $T_{sph}=131.7$ GeV~\cite{DOnofrio:2014rug}, which is given by $Y_B=\frac{28}{79}Y_{B-L}$ \cite{Pilaftsis:2003gt}.
Alternatively, we focus on the strong wash-out regime that $\tilde{m}/m_*\gg 1$, 
where $\tilde{m}$ is the effective neutrino mass defined by $v^2(yy^\dagger)/m_N$ 
and $m_* \approx 1.08 \times 10^{-3}$ eV is the equilibrium neutrino mass. 
Therefore, flavor effects in the charged-lepton sector are not expected to have a major influence on our results~\cite{Hambye:2016sby}.

\section{Dark matter}
\label{sec:4}

\begin{figure}[h]
\centering
\includegraphics[width=9.5cm,height=6.5cm]{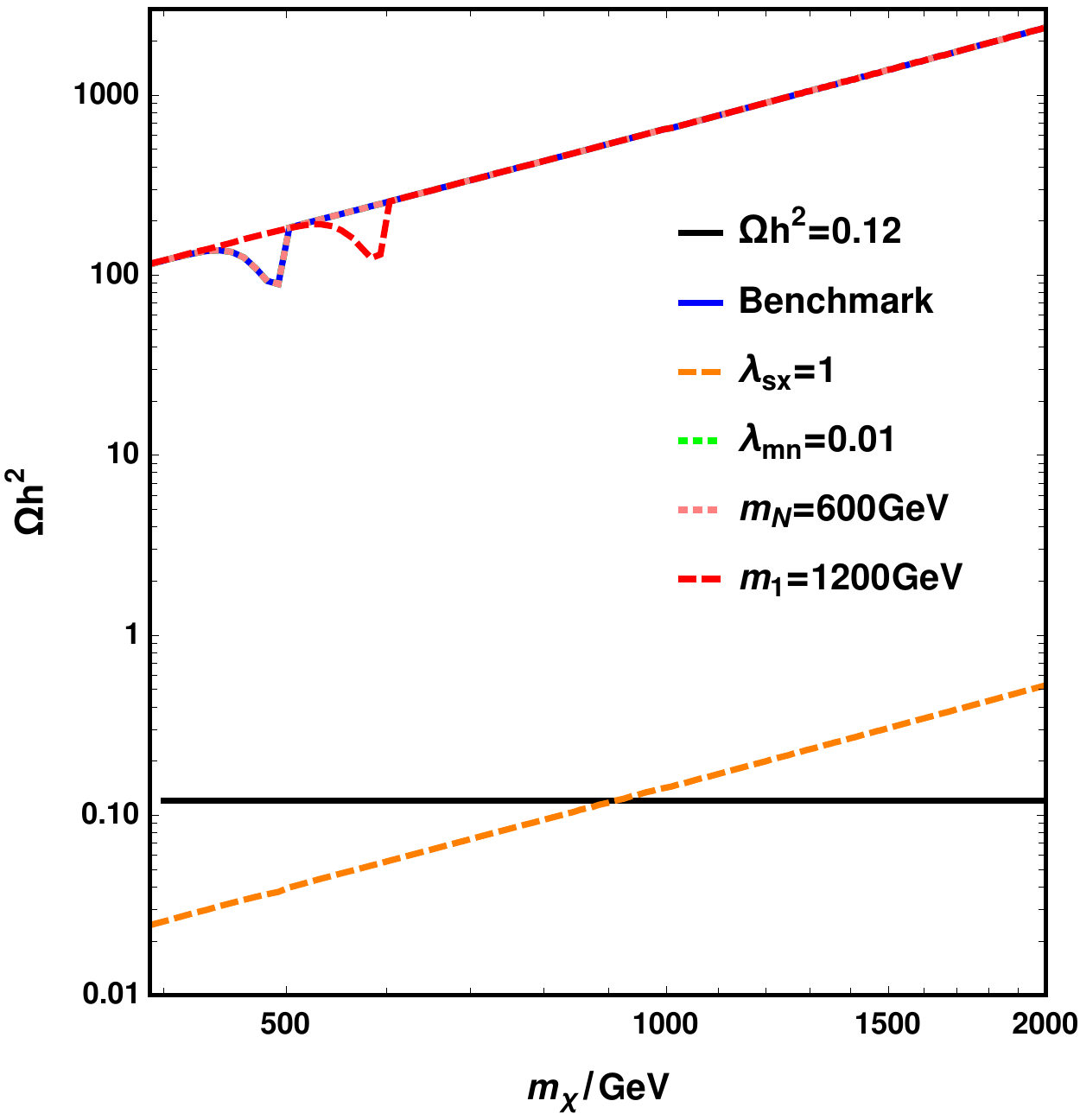}
\caption{Evolution of dark matter relic density with dark matter mass $m_{\chi}$, where the black line is the observed value with $\Omega h^2=0.12$ \cite{Planck:2015ica} where we fix $\sin\theta=0.01$. The blue line is the benchmark line we choose  $m_2=1500$ GeV,  $m_1=1000$ GeV, $\lambda_{sx}=0.1$, $\lambda_{mn}=0.1$ and $m_N=800$ GeV, and other colored lines correspond to the case varying one of the parameters.}
\label{Fig:fig2}
\end{figure}

\begin{figure}[h]
\centering
\includegraphics[width=9.5cm,height=6.5cm]{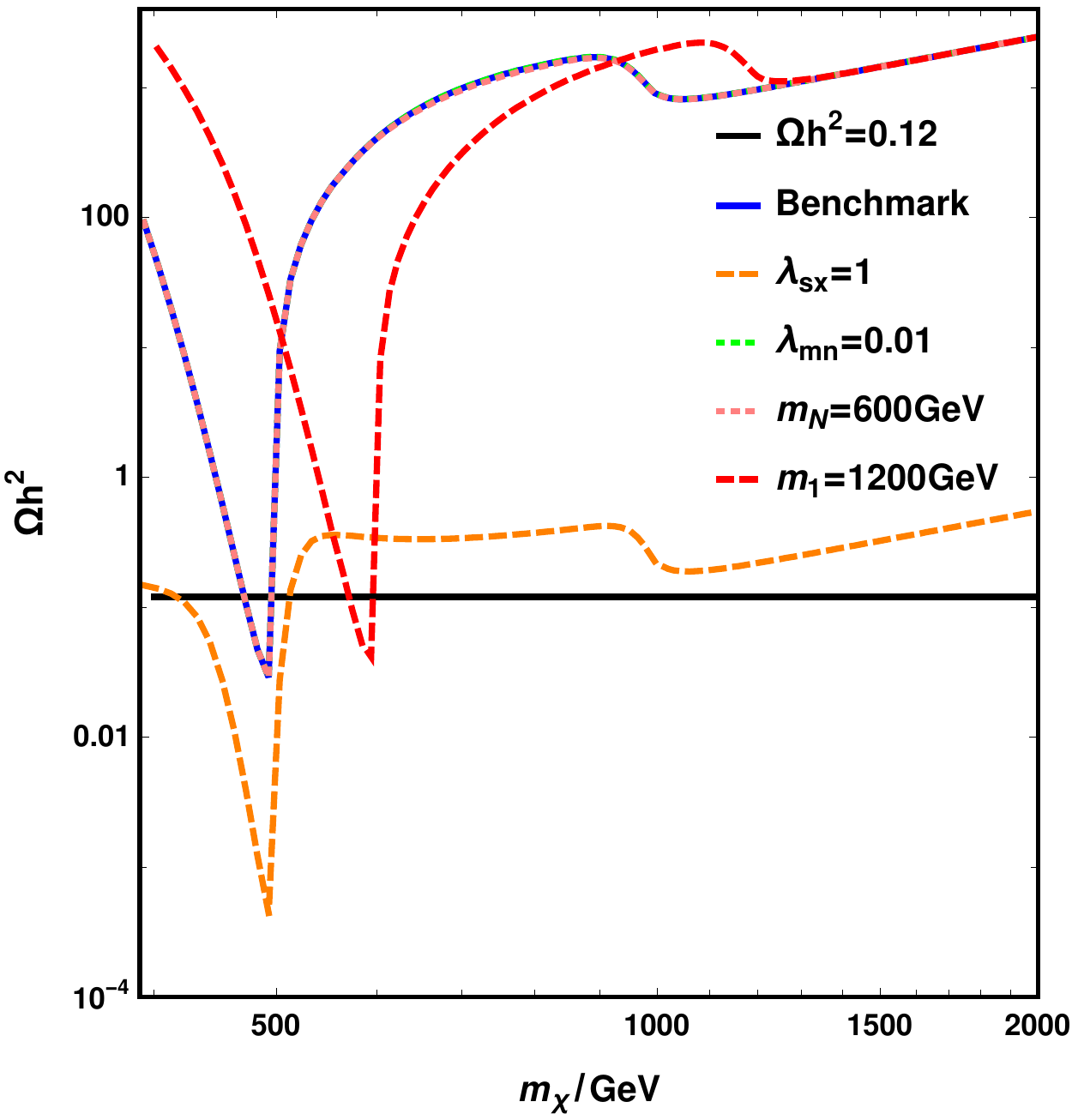}
\caption{Evolution of dark matter relic density with dark matter mass $m_{\chi}$, where the black line is the observed value with $\Omega h^2=0.12$ \cite{Planck:2015ica}where we fix $\sin\theta=0.9$. The blue line is the benchmark line we choose  $m_2=1500$ GeV, $m_1=1000$ GeV, $\lambda_{sx}=0.1$, $\lambda_{mn}=0.1$ and $m_N=800$ GeV, and other colored lines correspond to the case varying one of the parameters.}
\label{Fig:fig3}
\end{figure}

Technically, we implement the model with  Feynrules \cite{Alloul:2013bka}, and calculate the relic density with MicrOMEGAs~\cite{Belanger:2013oya} numerically. 
We give the evolution of dark matter relic density with dark matter mass 
$m_{\chi}$ in Fig.~\ref{Fig:fig2} and Fig.~\ref{Fig:fig3} corresponding to $\sin\theta=0.01$ and $\sin\theta=0.9$ respectively, 
where dark matter mass is set within $[400$ GeV, 2000 GeV]. 
In both figures, the black line is the observed value with $\Omega h^2=0.12$ \cite{Planck:2015ica}. 
The blue line is the benchmark line we choose $m_2=1500$ GeV, $m_1=1000$ GeV,
$\lambda_{sx}=0.1$, $\lambda_{mn}=0.1$ and $m_N=800$ GeV, 
and other colored lines correspond to the case varying one of the parameters. 

In Fig.~\ref{Fig:fig2}, the small $\sin\theta$ value limits the contribution of right-handed neutrinos to dark matter, and new Higgses play a more important role in determining DM relic density. The relic density increases with the increase of $m_{\chi}$ but we have a peak in the case of $m_{\chi} \approx 1/2 m_1$ due to the s-channel resonant-enhanced processes of $\chi\chi \to NN$. However, such a resonant-enhanced effect does not decrease relic density a lot due to the small $\sin\theta$ as we mentioned above. On the other hand, the small dark matter-scalar Yukawa coupling $\lambda_{sx}$ induces a small annihilation cross section so that dark matter is over-abundant, and dark matter relic density much decreases in the case of $\lambda_{sx}=1$, where the processes of dark matter annihilation into new Higgses are dominant and the peak arising from s-channel enhanced resonant is not obvious. 

According to Fig.~\ref{Fig:fig3}, we have different results in the case of $\sin\theta =0.9$ since right-handed neutrinos can also play important role in determining dark matter relic density. We have a resonant region at about $m_{\chi}=1/2m_1$, where the relic density drops sharply, and intersect with the relic density constraint curve. What's more, we have another peak in the case of $m_{\chi} \approx m_1$, where the t-channel Higgs-mediated processes open so that decreasing the relic density. In both figures, the lines correspond to different $\lambda_{mn}$ and $m_N$ are almost coincide with the benchmark line, which indicates $m_N$ and small $\lambda_{mn}$ can make little difference on relic density.

Aside from the observed relic density constraint, direct detection for dark matter puts the most stringent limit on the dark matter parameter space. Concretely speaking, for dark matter mass $m_{\chi} \gtrsim 2$ GeV, the direct detection experiments give the most stringent
constraint on the spin-independent DM matter scattering with nucleon, 
 and XENON1T \cite{XENON:2020gfr} gives a stringent bound for $m_{\chi} >$ 6 GeV. On
the other hand, considering neutrino floor limit \cite{Billard:2021uyg,Billard:2013qya} on the current probe sensitive to dark matter, WIMP dark matter is facing a serious crisis since there is no evidence for the existence of dark matter. Fortunately, in our work, direct detection constraint is much weak since processes involving  dark matter are new Higgses as well as right-handed neutrinos but SM particles are almost irrelevant.

\section{Discussion}
\label{sec:5}

\subsection{Part I}
   \label{part1}
In this section, we discuss the interplay between dark matter and leptogenesis in our framework. 
Firstly, we discuss the parameter space related to dark matter. There are seven parameters in our model with 
  \begin{eqnarray}
  m_{\chi}, m_1, m_2, \sin\theta , \lambda_{sx}, \lambda_{mn}, m_N.
  \end{eqnarray}
For different $\sin\theta$, we will have different parameter space depending on the contribution of right-handed neutrino to relic density. 
Therefore, we consider two cases of $\sin\theta=0.01$ and $\sin\theta=0.9$, 
where right-handed neutrinos can play an important role in determining dark matter relic density in the latter case. 
For $m_1$ and $m_2$, we consider two cases with $m_2=1.5m_1$ and $m_2=0.8m_1$ for simplicity. 
Particlularly, we will come to the Forbidden-DM case when $m_{\chi}<m_{1,2}$, 
and we assume $ m_{1,2}<2m_{\chi}$ to avoid the case that the production $h_{1,2}$ decay into a pair of $\chi$ for simplicity, 
while the scalars that are produced in DM annihilation can subsequently decay to light neutrinos due to the heavy-light neutrino mixing.
We scan the parameter space with,
 \begin{eqnarray} \nonumber
 &&m_{\chi} \subset [300\ \mathrm{GeV},1500\ \mathrm {GeV}], \\\nonumber
 &&m_1 \subset [900\ \mathrm{ GeV}, 1500\ \mathrm {GeV}], \\\nonumber
 &&m_N \subset[600\ \mathrm{GeV},1300\ \mathrm{GeV}], \\\nonumber
 &&\lambda_{sx} \subset [0.01,\sqrt{4\pi}], \\
 &&\lambda_{mn} \subset [0.001,0.1].
 \end{eqnarray} 
We give the results satisfying dark matter relic density constraint in the following figures, where Fig.~\ref{Fig:4} to Fig.~\ref{Fig:7} correspond to $m_2=1.5m_1$, and Fig.~\ref{Fig:8} to Fig.~\ref{Fig:11} is $m_2=0.8m_1$. 

In Fig.~\ref{Fig:4} and Fig.~\ref{Fig:5}, we have $\sin\theta=0.01$ and $m_2=1.5m_1$, and the dominant processes related to dark matter relic density are $\chi\chi\to h_{1,2}h_{1,2}$. The viable region for $\lambda_{sx}$ satisfying relic density constraint is limited  at $O(1)$ level. For a smaller $\lambda_{sx}$, the DM annihilation cross section can be so small that DM will be over-abundant and for the larger $\lambda_{sx}$, we will have dark matter under-abundant due to the large cross section. On the other hand, $\lambda_{sx}$ is also related to dark matter mass, and a larger $\lambda_{sx}$ always corresponds to a larger $m_{\chi}$ as we can see from Fig.~\ref{Fig:4}. As we mentioned above, $\lambda_{mn}$  makes little difference in the dark matter relic density in this case that one can always obtain the correct relic density among the chosen parameter space with $0.001 \leqslant \lambda_{mn}\leqslant 0.1$. In Fig.~\ref{Fig:5}, we give the result of the $m_N-m_{\chi}$ satisfying DM relic density constraint, where points with different colors represent $m_1$ taking different values. Since the contribution of the right-handed neutrino to relic density is limited like $\lambda_{mn}$ in this case, we also have a flexible region for $m_N$ value. For $m_{\chi}<600$ GeV, the process of $\chi\chi\to h_{1,2}$ is suppressed due to the large relative mass splitting of $m_{\chi}$ with $m_{1}$ \cite{DAgnolo:2015ujb}, and such region is excluded for the over-abundant dark matter.

In Fig.~\ref{Fig:6} and Fig.~\ref{Fig:7}, we give the results in the case of $\sin\theta=0.9$, where annihilation of dark matter into right-handed neutrinos will also make difference on the relic density, and we have a wider parameter space for $\lambda_{sx}$ with $1<\lambda_{sx}<1.8$ as we can see in Fig.~\ref{Fig:6}. On the other hand, $m_{\chi}$ will not just increase with the increase of $\lambda_{sx}$ like the case of $\sin\theta=0.01$ because of the extra processes. In Fig.~\ref{Fig:7}, we give the result of the $m_N-m_{\chi}$ satisfying DM relic density constraint, where points with different colors represent $m_1$ taking different values.
Notice that we will also come to the Forbidden-DM scenario with $\chi\chi \to NN$ in the case of $m_{\chi}<m_N$, 
and we can have the similar result with $\sin\theta=0.01$ when the relative mass splitting of $m_{\chi}$ with $m_N$ is large so that $\chi\chi \to NN$ is highly suppressed. 

 
In the case of $m_1>m_2$  we set $m_2=0.8m_1$. Due to the small mass splitthing of $m_{\chi}$ with $m_2$, dark matter mass can decrease to 500 GeV with the correct relic density as in Fig.~\ref{Fig:9} and Fig.~\ref{Fig:11}. On the other hand, we will have a wider parameter space for $\lambda_{mn}-\lambda_{sx}$ in the case of $\sin\theta=0.01$ and $\sin\theta=0.9$.
  We give the result of $m_N-m_{\chi}$ satisfying dark matter constraint in Fig.~\ref{Fig:9} and Fig.~\ref{Fig:11} corresponding to the case of $\sin\theta=0.01$ and $\sin\theta=0.9$ respectively, where points with different colors represent $m_2$ taking different values. Similarly, we have a more flexible parameter space for $m_N-m_{\chi}$ for the light $m_2$.
 \begin{figure}[h]
\centering
\begin{minipage}[t]{0.48\textwidth}
\centering
\includegraphics[width=7cm,height=5cm]{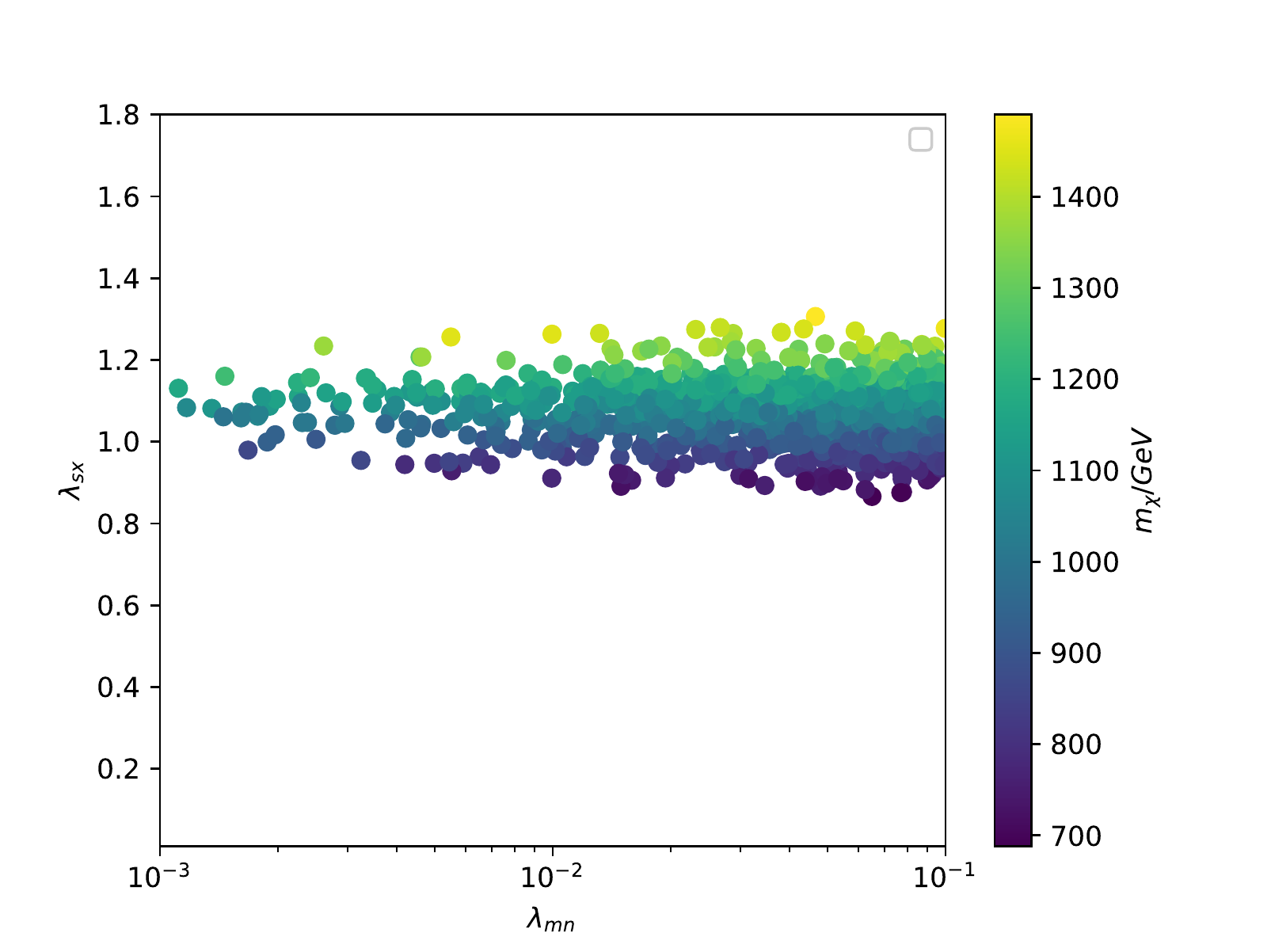}
\caption{Results of the $\lambda_{mn}-\lambda_{sx}$ satisfying dark matter relic density constraint in the case of $sin\theta=0.01$ and $m_2=1.5m_1$, where points with different color correspond to $m_{\chi}$ taking different value.}
  \label{Fig:4}
\end{minipage}
\begin{minipage}[t]{0.48\textwidth}
\centering
\includegraphics[width=7cm,height=5cm]{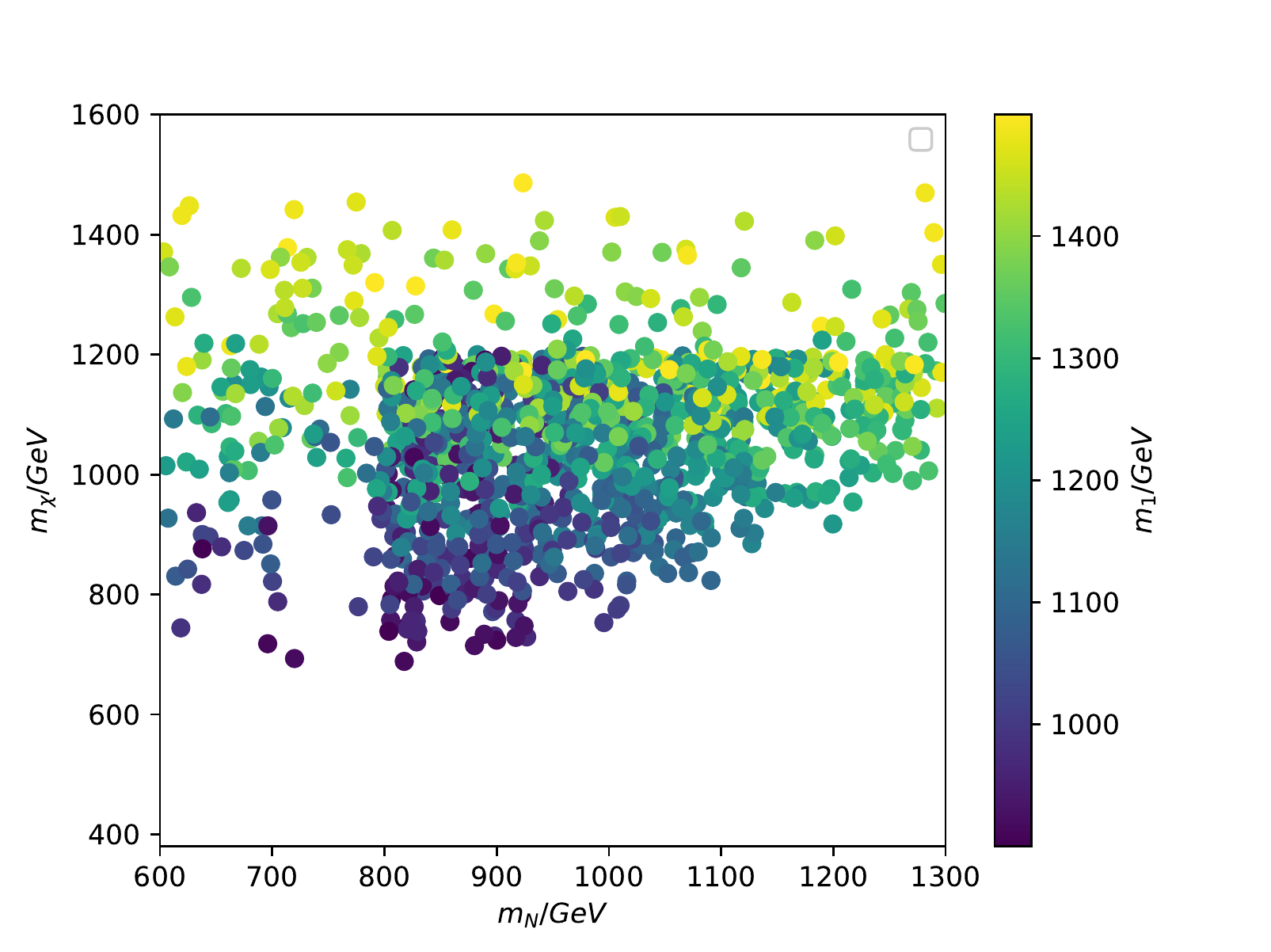}
\caption{Results of the $m_N-m_{\chi}$ satisfying dark matter relic density constraint in the case of $sin\theta=0.01$ and $m_2=1.5m_1$, where points with different color correspond to $m_1$ taking different value.}
 \label{Fig:5}
\end{minipage}
\end{figure}
\begin{figure}[h]
\begin{minipage}[t]{0.48\textwidth}
\centering
\includegraphics[width=7cm,height=5cm]{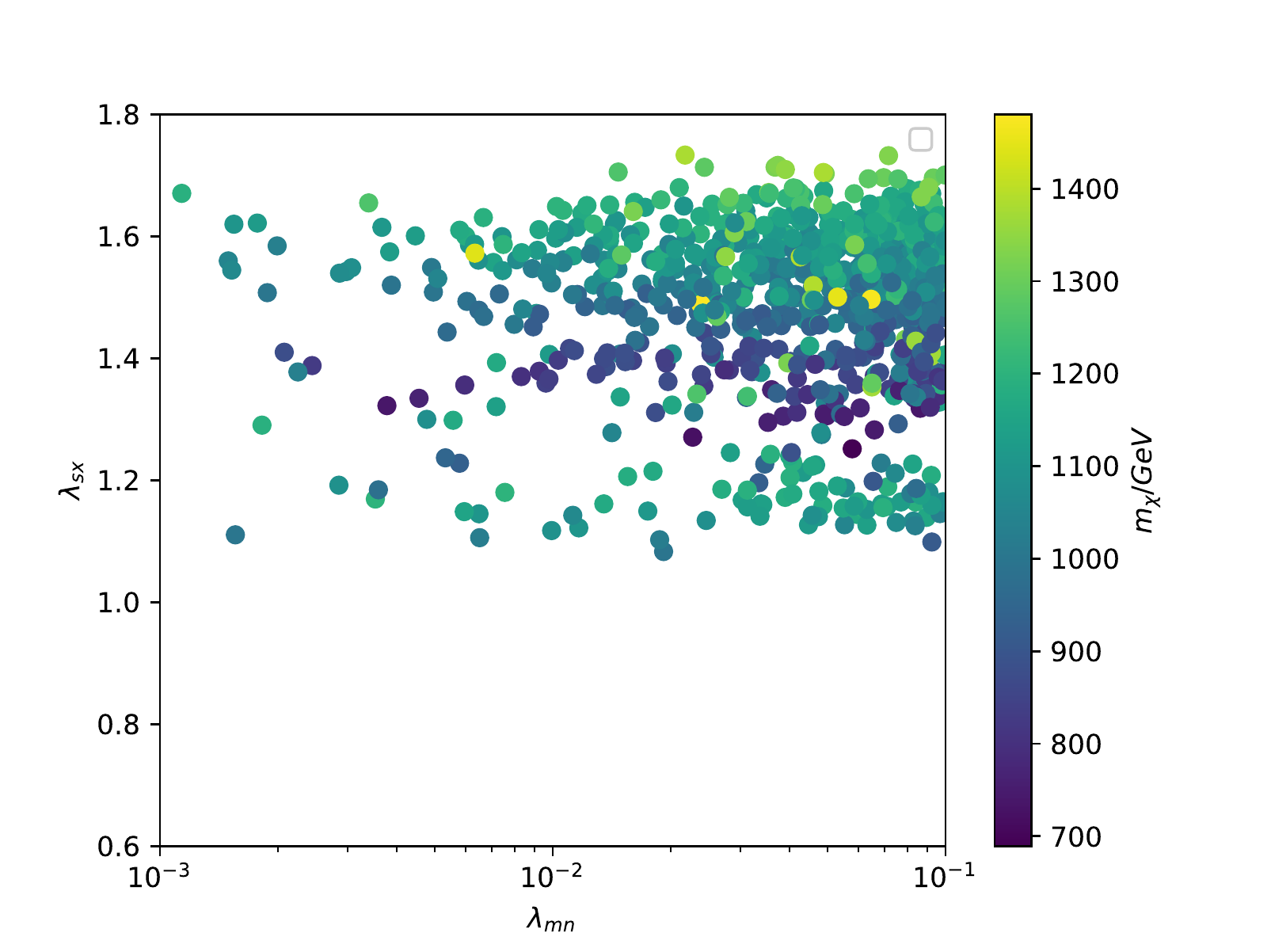}
\caption{Results of the $\lambda_{mn}-\lambda_{sx}$ satisfying dark matter relic density constraint in the case of $sin\theta=0.9$ and $m_2=1.5m_1$, where points with different color correspond to $m_{\chi}$ taking different value.}
  \label{Fig:6}
\end{minipage}
\centering
\begin{minipage}[t]{0.48\textwidth}
\centering
\includegraphics[width=7cm,height=5cm]{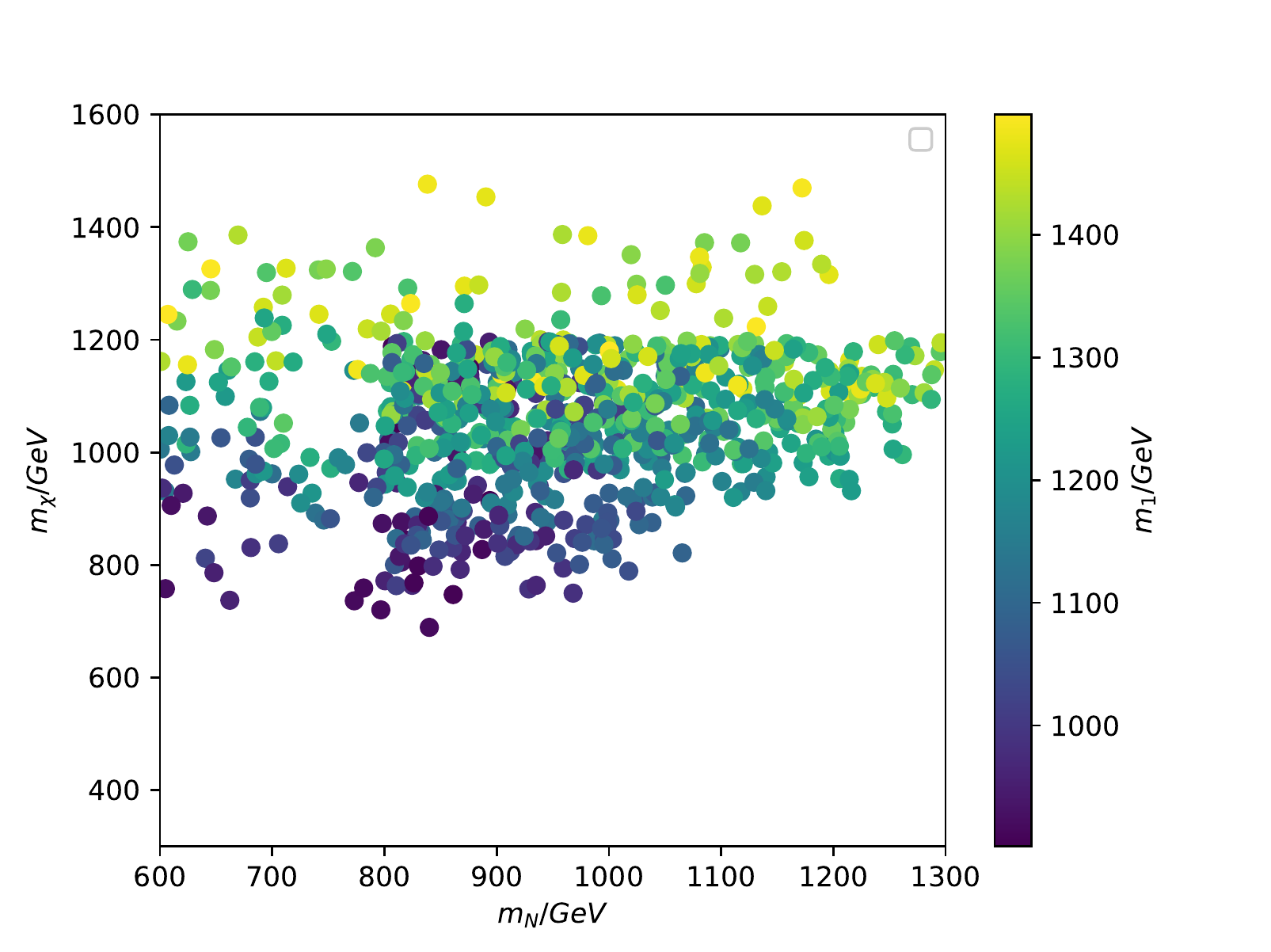}
\caption{Results of the $m_N-m_{\chi}$ satisfying dark matter relic density constraint in the case of $sin\theta=0.9$ and $m_2=1.5m_1$, where points with different color correspond to $m_1$ taking different value.}
 \label{Fig:7}
\end{minipage}
\end{figure}
\begin{figure}[h]
\begin{minipage}[t]{0.48\textwidth}
\centering
\includegraphics[width=7cm,height=5cm]{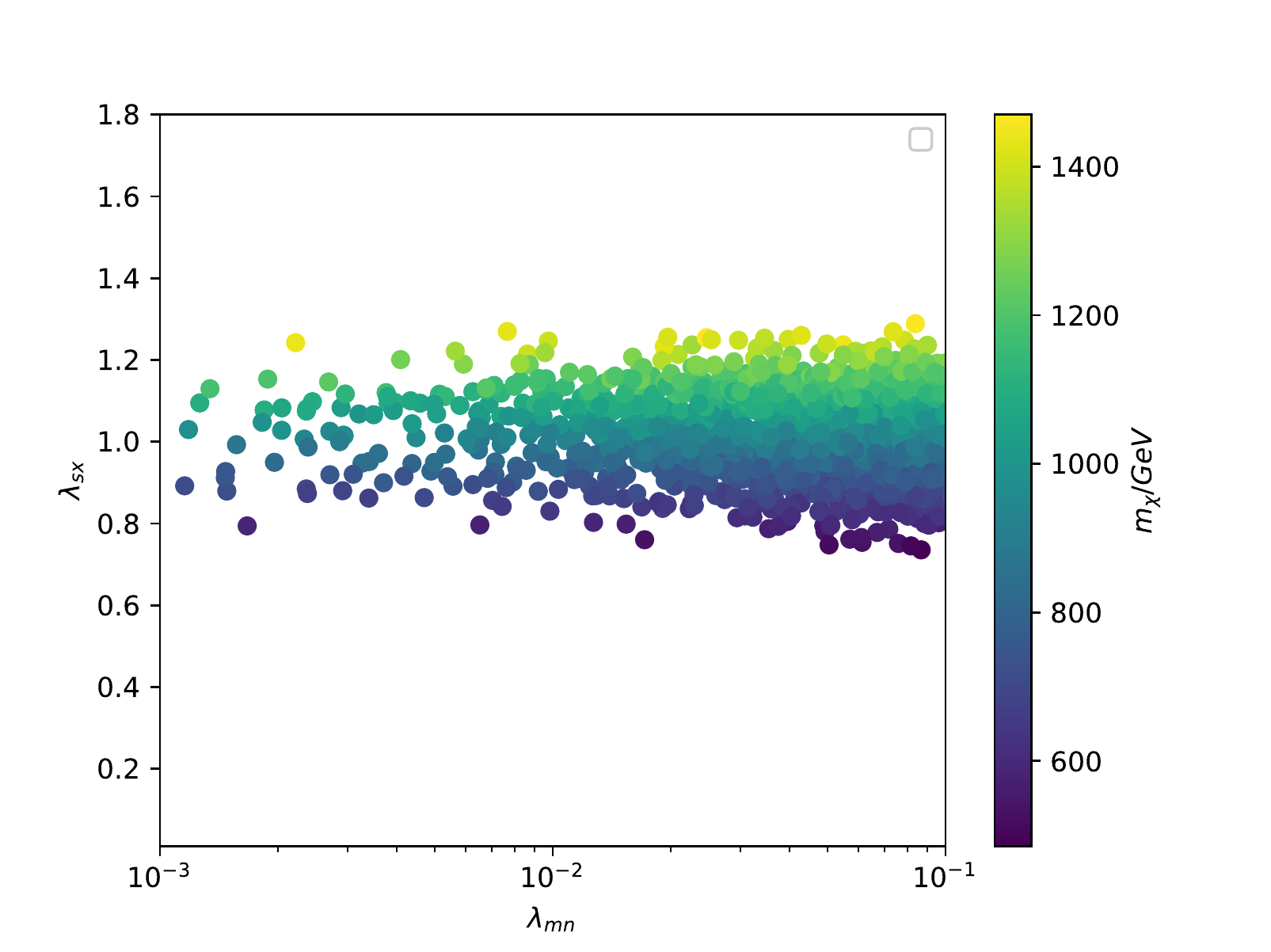}
\caption{Results of the $\lambda_{mn}-\lambda_{sx}$ satisfying dark matter relic density constraint in the case of $sin\theta=0.01$ and $m_2=0.8m_1$, where points with different colors correspond to $m_{\chi}$ taking different value.}
 \label{Fig:8}
\end{minipage}
\centering
\begin{minipage}[t]{0.48\textwidth}
\centering
\includegraphics[width=7cm,height=5cm]{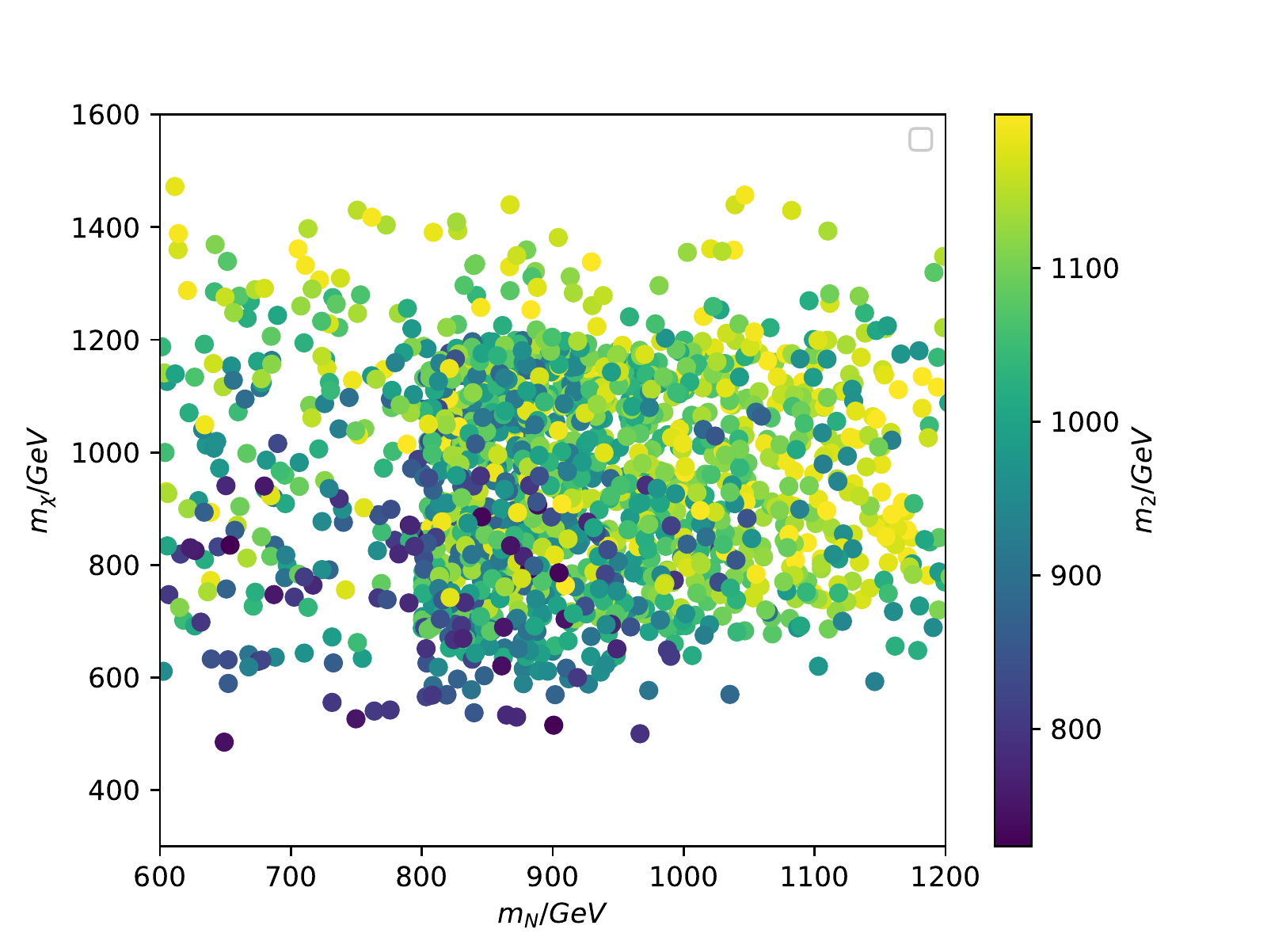}
\caption{Results of the $m_N-m_{\chi}$ satisfying dark matter relic density constraint in the case of $sin\theta=0.01$ and $m_2=0.8m_1$, where points with different colors correspond to $m_2$ taking different value.}
 \label{Fig:9}
\end{minipage}
\end{figure}
\begin{figure}[h]
\begin{minipage}[t]{0.48\textwidth}
\centering
\includegraphics[width=7cm,height=5cm]{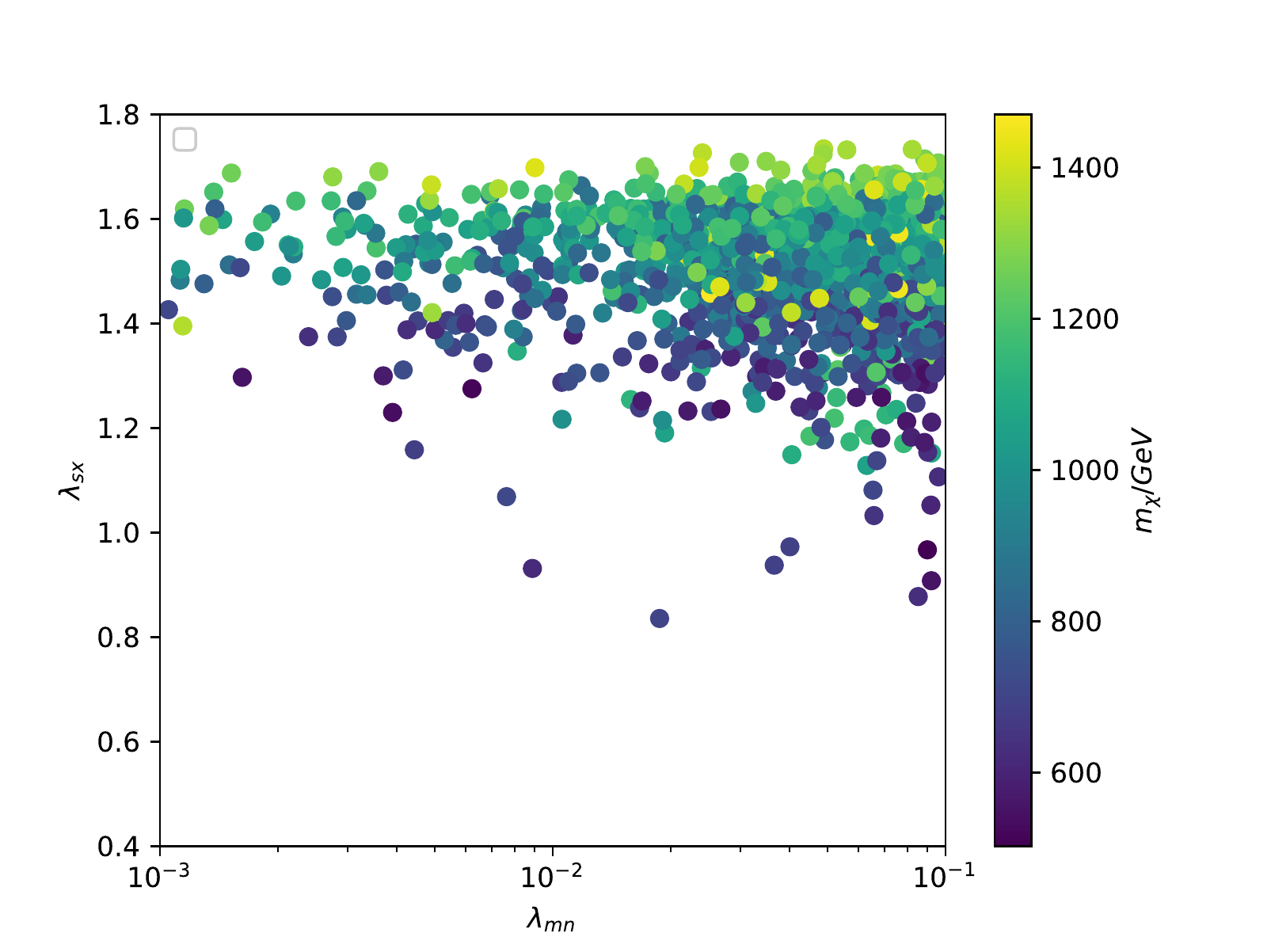}
\caption{Results of the $\lambda_{mn}-\lambda_{sx}$ satisfying dark matter relic density constraint in the case of $sin\theta=0.9$ and $m_2=0.8m_1$, where points with different colors correspond to $m_{\chi}$ taking different value.}
 \label{Fig:10}
\end{minipage}
\centering
\begin{minipage}[t]{0.48\textwidth}
\centering
\includegraphics[width=7cm,height=5cm]{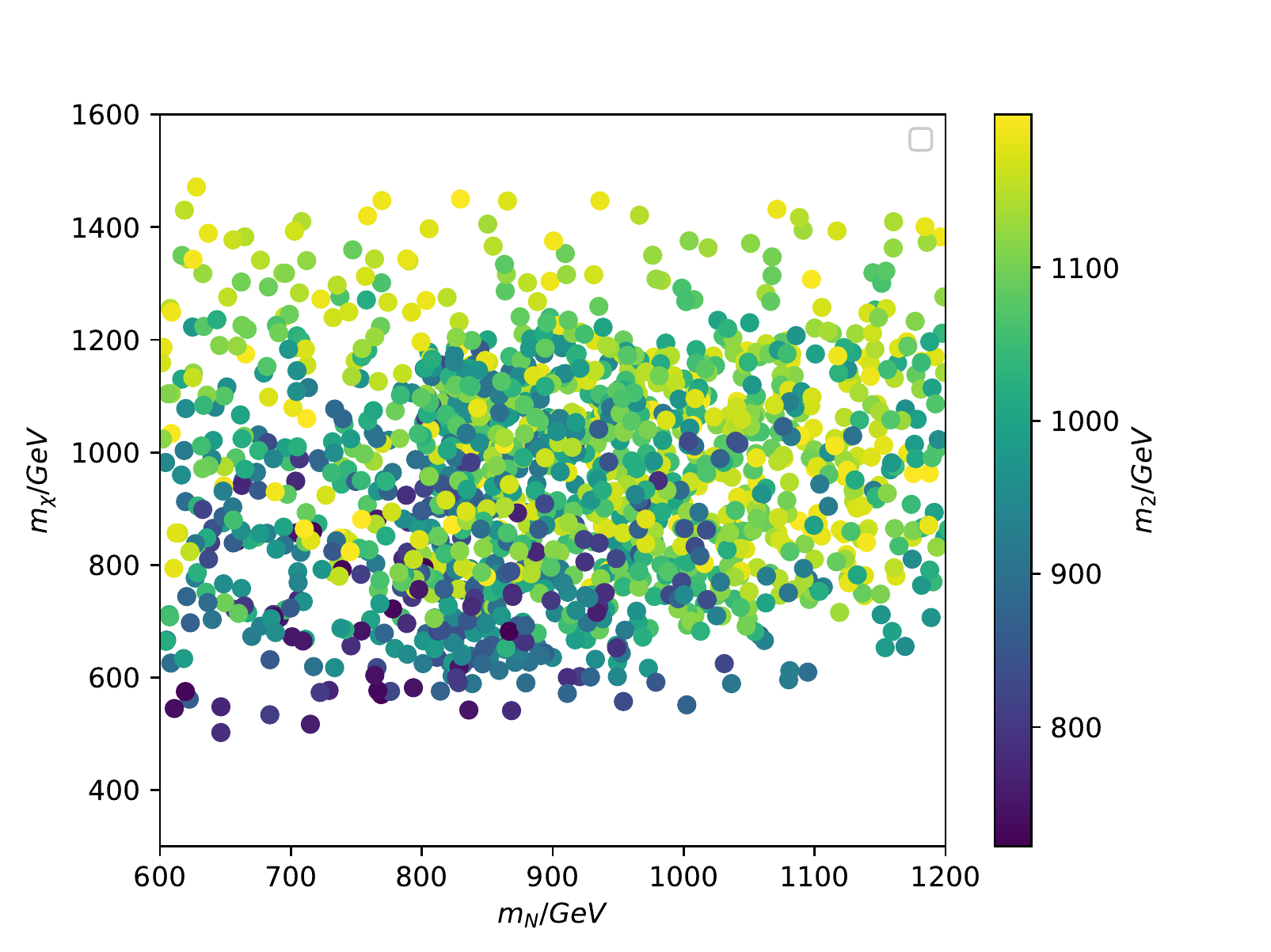}
\caption{Results of the $m_N-m_{\chi}$ satisfying dark matter relic density constraint in the case of $sin\theta=0.9$ and $m_2=0.8m_1$, where points with different colors correspond to $m_2$ taking different value.}
 \label{Fig:11}
\end{minipage}
\end{figure}

\subsection{Part II}
    \label{part2}
 \begin{figure}[h]
\begin{minipage}[t]{0.48\textwidth}
\centering
\includegraphics[width=7cm,height=5cm]{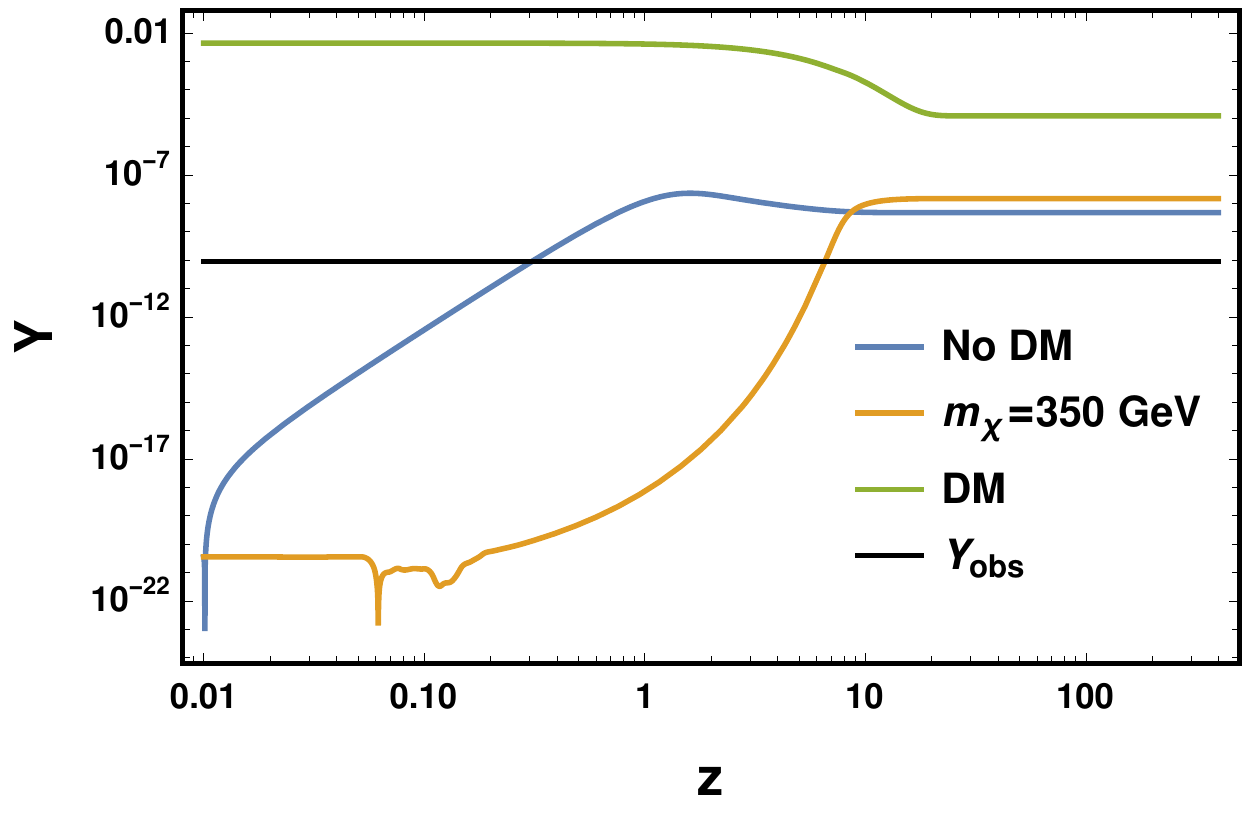}
\caption{
Evolution of the baryon asymmetry  as well as dark matter abundance in the case of  $m_2=1.5m_1$, $m_N=600$ GeV and $m_{\chi}=350$ GeV, where the black line represents the observed baryon asymmetry, the green and orange lines correspond to the results of the baryon asymmetry as well as dark matter abundance, while the blue line represents the evolution of baryon asymmetry without dark matter in the model.}
 \label{Fig:8a}
\end{minipage}
\centering
\begin{minipage}[t]{0.48\textwidth}
\centering
\includegraphics[width=7cm,height=5cm]{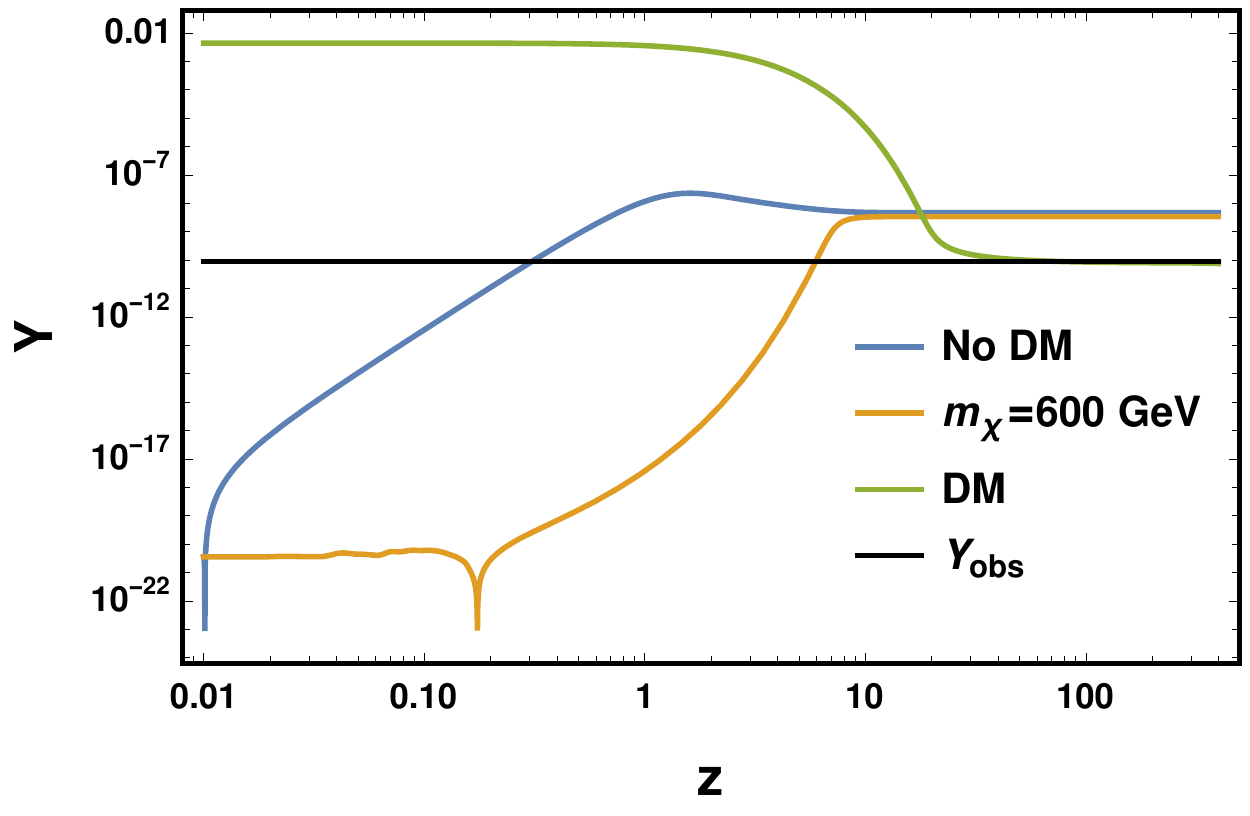}
\caption{Evolution of the baryon asymmetry as well as dark matter abundance in the case of  $m_2=1.5m_1$, $m_N=600$ GeV and $m_{\chi}=600$ GeV, where the black line represents the observed baryon asymmetry, the green and orange lines correspond to the results of the baryon asymmetry as well as dark matter abundance, while the blue line represents the evolution of baryon asymmetry without dark matter in the model.}
 \label{Fig:9a}
\end{minipage}
\end{figure}
\begin{figure}[h]
\begin{minipage}[t]{0.48\textwidth}
\centering
\includegraphics[width=7cm,height=5cm]{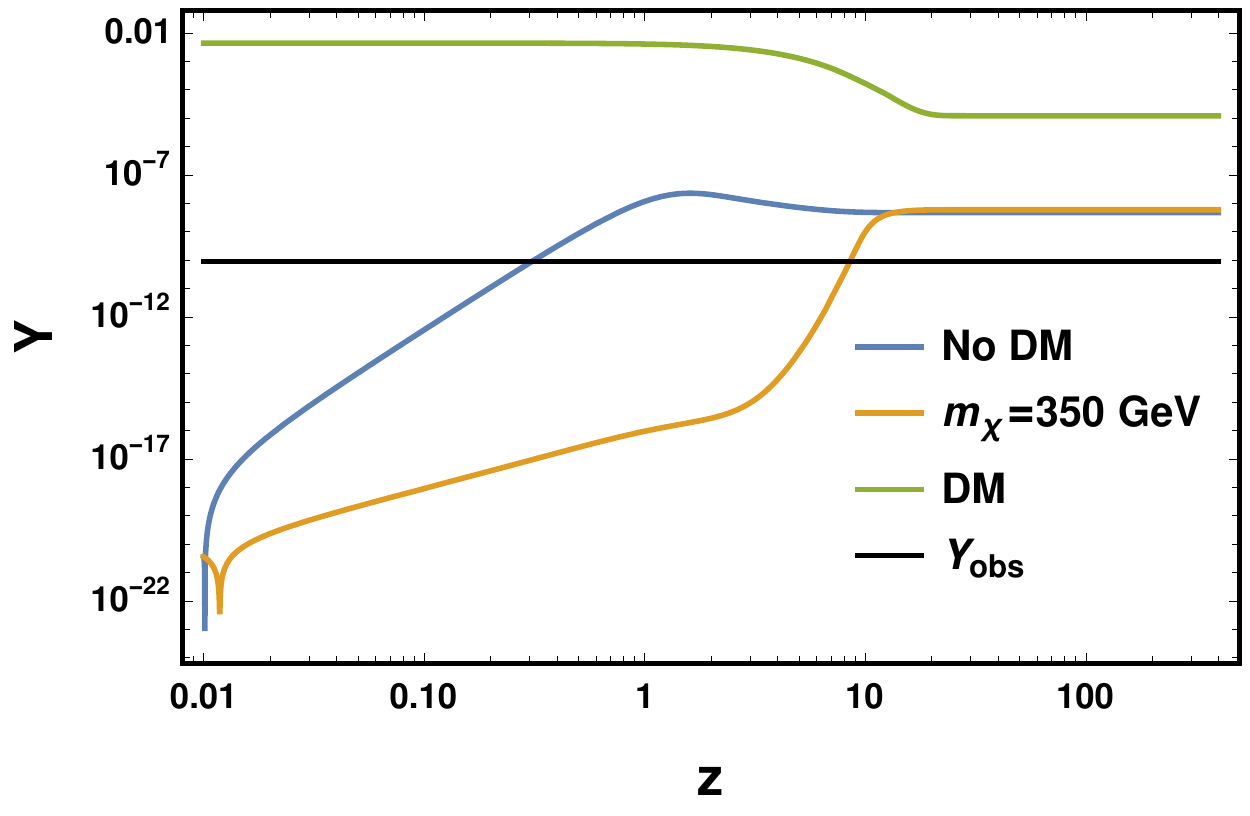}
\caption{Evolution of the baryon asymmetry  as well as dark matter abundance in the case of  $m_2=0.8m_1$, $m_N=600$ GeV and $m_{\chi}=350$ GeV, where the black line represents the observed baryon asymmetry, the green and orange lines correspond to the results of the baryon asymmetry as well as dark matter abundance, while the blue line represents the evolution of baryon asymmetry without dark matter in the model.}
 \label{Fig:10a}
\end{minipage}
\centering
\begin{minipage}[t]{0.48\textwidth}
\centering
\includegraphics[width=7cm,height=5cm]{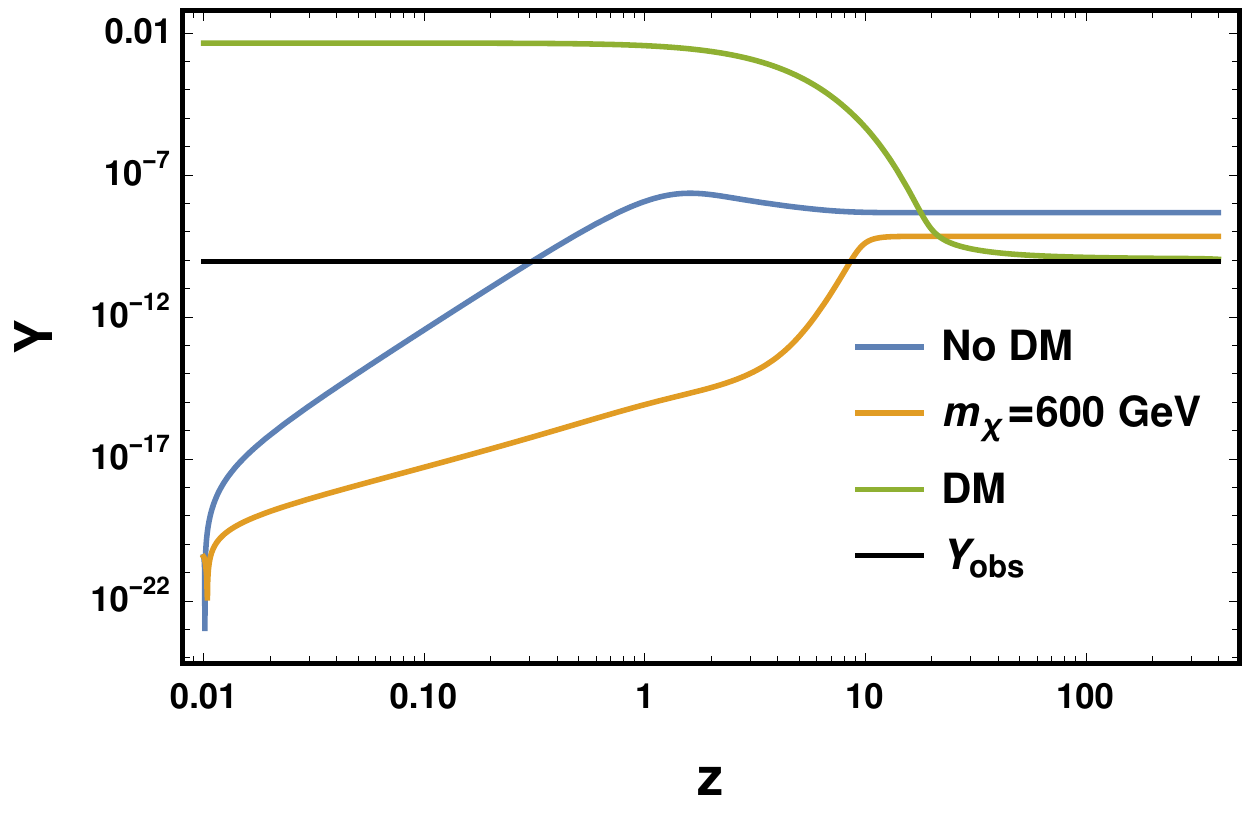}
\caption{Evolution of the baryon asymmetry as well as dark matter abundance in the case of  $m_2=0.8m_1$, $m_N=600$ GeV and $m_{\chi}=600$ GeV, where the black line represents the observed baryon asymmetry, the green and orange lines correspond to the results of the baryon asymmetry as well as dark matter abundance, while the blue line represents the evolution of baryon asymmetry without dark matter in the model.}
 \label{Fig:11a}
\end{minipage}
\end{figure}

 The interplay between dark matter and leptogenesis is determined by the  process of $\chi\chi \to NN$ and  the inverse process $NN \to \chi\chi$ together. These two processes are not always in thermal equilibrium since the number density of either $N$ or $\chi$ can be highly exponent suppressed when the temperature below the mass of $N$($\chi$).
 For the process of $\chi\chi \to NN$, more right-handed neutrinos that can decay have been generated and the baryon asymmetry is enhanced while for the inverse process right-handed neutrinos annihilate into the dark matter which dilutes the baryon asymmetry. 
 
Notice that the cross section of such a process is proportional to $\lambda_{sx}^2\lambda_{mn}^2\sin^2\theta$, 
and dark matter as well as baryon asymmetry will be less relevant due to the smaller $\sin\theta$, $\lambda_{sx} $ and $\lambda_{mn}$. 
In addition, according to the above discussion, the parameter space of $m_{\chi}-\lambda_{sx}$ is well constrained in the case of $\sin\theta=0.01$. 
Therefore, we consider the case of $\sin\theta=0.9$ in order to discuss the interplay between dark matter and leptogenesis. 
We fix $\lambda_{sx}=1.3$, $m_1=1200$ GeV and $\lambda_{mn}=0.1$, 
which correspond to a flexible dark matter mass region satisfying relic density constraint. 
 
We evolve the Boltzmann equations in the case of $m_2=1.5m_1$ and $m_2=0.8m_1$ respectively with Eq.~\ref{eq4}, where $\epsilon_{CP}$ is set to be $10^{-4}$. To obtain the baryon asymmetry, we also fix $\tilde{m}=0.01$ eV so that the related Yukawa couplings can be given by $y=\frac{\sqrt{2m_N\tilde{m}}}{v}$. In Fig.~\ref{Fig:8a} and Fig.~\ref{Fig:9a}, we give the results of $m_2=1.5m_1$ while  Fig.~\ref{Fig:10a} and Fig.~\ref{Fig:11a} are the results of $m_2=0.8m_1$. The black lines represent the observed BAU $Y_B=8.6\times 10^{-11}$ \cite{Planck:2018vyg} while the orange(blue) lines correspond to the results of baryon asymmetry with(without) dark matter where we have fixed $m_N=600$ GeV, and the green lines represent the evolution of dark matter abundance. In Fig.~\ref{Fig:8a}, we set $m_{\chi}=350$ GeV, it is obvious the orange line is above the blue line which means the baryon asymmetry is enhanced distinctly since more right-handed neutrinos have been generated via the process of $\chi\chi \to NN$. On the other hand, in Fig.~\ref{Fig:9a} with $m_{\chi}=600$ GeV, the blue line is slightly above the orange line, indicating that the baryon asymmetry is diluted by the process of $NN \to \chi\chi$. Similar results can be found in Fig.~\ref{Fig:10a} and Fig.~\ref{Fig:11a} with $m_2=0.8m_1$, where the BAU is slightly enhanced for $m_{\chi} =350$ GeV and  significantly diluted in the case of  $m_{\chi} =600$ GeV.

\begin{figure}[h]
\centering
\includegraphics[width=9.5cm,height=6.5cm]{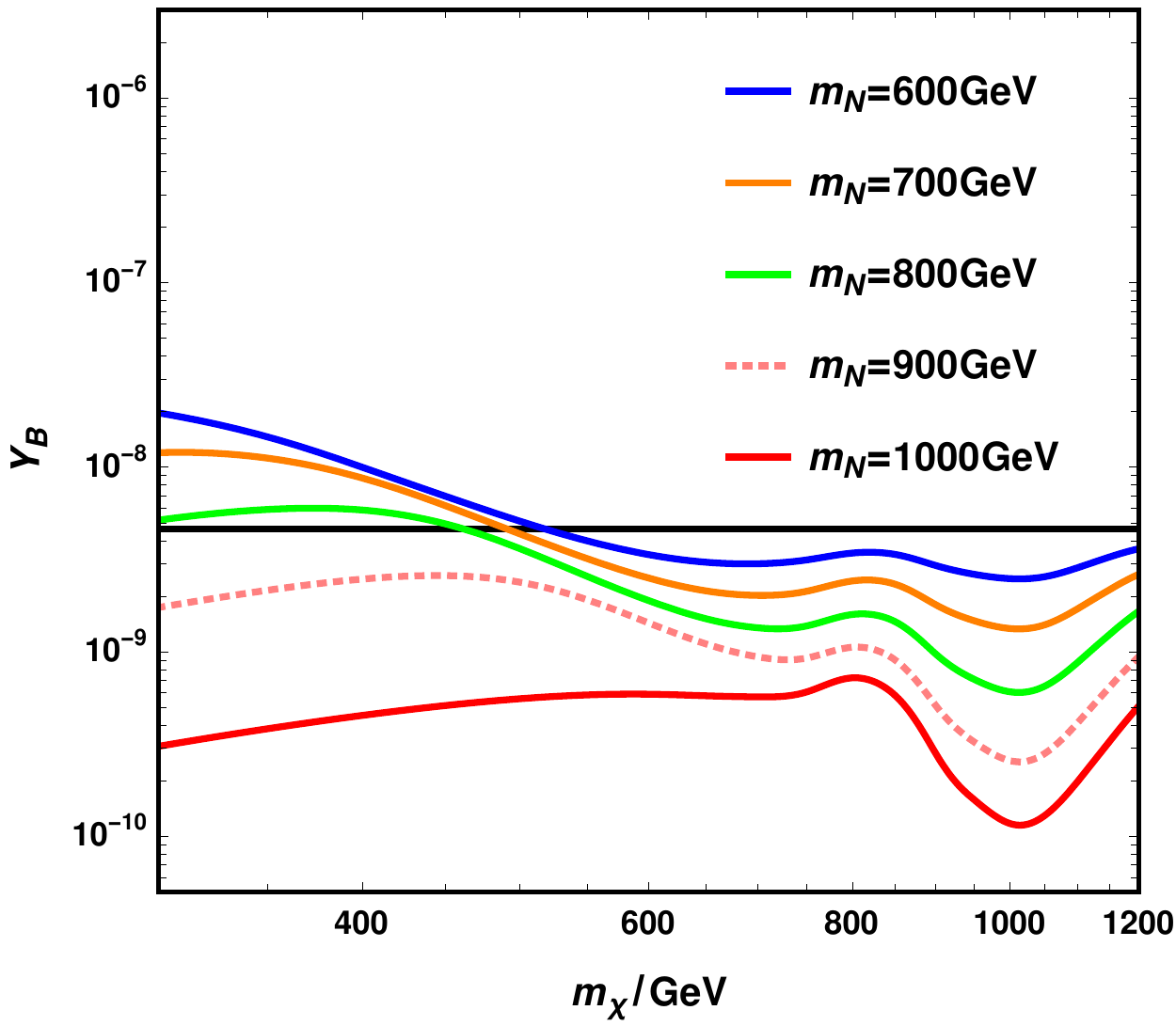}
\caption{
Relationship between baryon asymmetry $Y_B$ and dark matter mass $m_{\chi}$, 
where we fixed $m_1=1200$ GeV, $m_2=1.5m_1$, $\sin\theta=0.9$, $\lambda_{sx}=1.3$, $\lambda_{mn}=0.1$, 
the black line is the result without dark matter, while other colored lines correspond to the case that $m_N$ takes value from 600 GeV to 1000 GeV.}
\label{Fig:12}
\end{figure}

 Furthermore, in Fig.~\ref{Fig:12}, we give the relationship between the baryon asymmetry $Y_B$ and dark matter mass in the case of $m_2=1.5m_1$, 
where dark matter mass is constrained within $[300$ GeV, 1200 GeV]. For the heavier dark matter, dark matter can make little difference in the baryon asymmetry 
since the heavier dark matter may be frozen-out while right-handed neutrinos are still in thermal equilibrium. 
The black line corresponds to the baryon asymmetry value without dark matter when evaluating Boltzmann equations, 
and other colored lines correspond to $m_N$ taking values from 600 GeV to 1000 GeV. 
Note that the region above the black line represents the baryon asymmetry is strengthed while the leptogenesis result is diluted below the black line. 

For $m_N=600, 700$  and 800 GeV, the left part of the curves corresponding to 
the light dark matter mass region are above the black line, which indicates the baryon asymmetry is strengthed.  
As dark matter mass becomes larger, the dilution effect becomes efficient so that the generated baryon asymmetry is smaller than the case without dark matter.
Particularly, in the case of $m_{\chi}=m_2/2$, 
the baryon asymmetry dilution becomes most obvious corresponding to the peak of the curve due to the resonance-enhanced.
In the case of $m_N=900, 1000$ GeV, 
 the corresponding curves are both below the black line, which means the baryon asymmetry is exactly diluted by $NN \to \chi\chi$.

\begin{figure}[h]
\centering
\includegraphics[width=9.5cm,height=6.5cm]{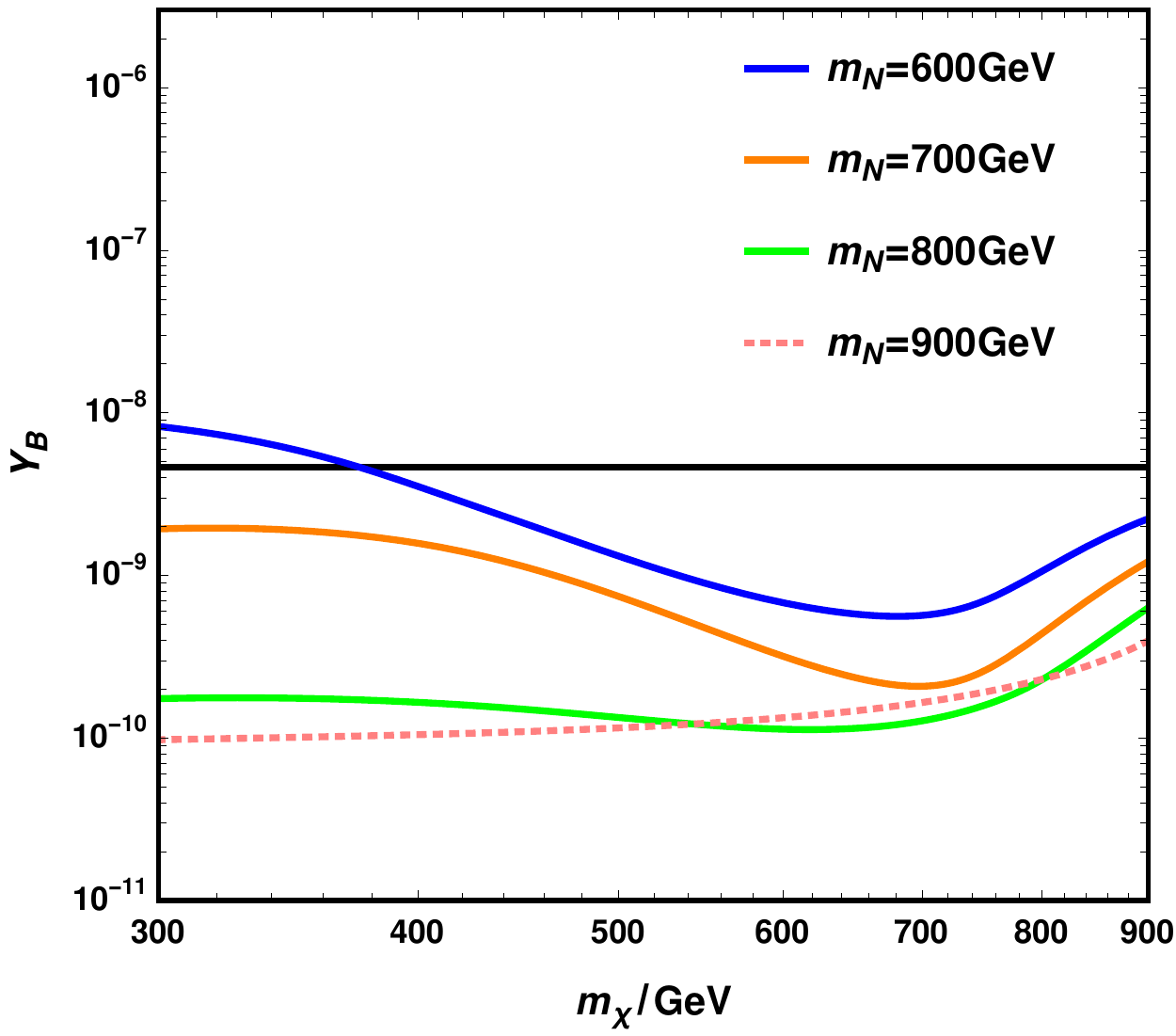}
\caption{
Relationship between baryon asymmetry $Y_B$ and dark matter mass $m_{\chi}$, 
where we fixed $m_1=1200$ GeV, $m_2=0.8m_1$, $\sin\theta=0.9$, $\lambda_{sx}=1.3$, $\lambda_{mn}=0.1$, 
the black line is the result without dark matter, while other colored lines correspond to the case that $m_N$ takes value from 600 GeV to 900 GeV.}
\label{Fig:13}
\end{figure}

In Fig.~\ref{Fig:13}, we give the relationship between baryon asymmetry $Y_B$ and dark matter mass in the case of $m_2=0.8m_1$, but $m_N$ takes values from 600 GeV to 900 GeV and dark matter mass is constrained with [300 GeV, 950 GeV]. For $m_N=600$ GeV, we still have baryon asymmetry-enhanced effect for light dark matter. However, for $m_N>600$ GeV, compared with the case of $m_2=1.5m_1$, the curves are all below the black line, 
which means baryon asymmetry is diluted with the process $NN \to \chi\chi$ induced by the lighter $m_2$.
  
It is worth stressing that with the increase of dark matter mass, 
the baryon asymmetry will be approximate to the value without dark matter case until the result is almost not affected 
by the existence of dark matter, which corresponds to the case that dark matter freeze-out is much earlier than leptogenesis freeze-out. 
Similarly, for the smaller dark matter mass, dark matter will also make little difference in the baryon asymmetry 
since leptogenesis freeze-out is much earlier than dark matter freeze-out.

\subsection{Part III}

\begin{figure}[h]
\begin{minipage}[t]{0.48\textwidth}
\centering
\includegraphics[width=6cm,height=6cm]{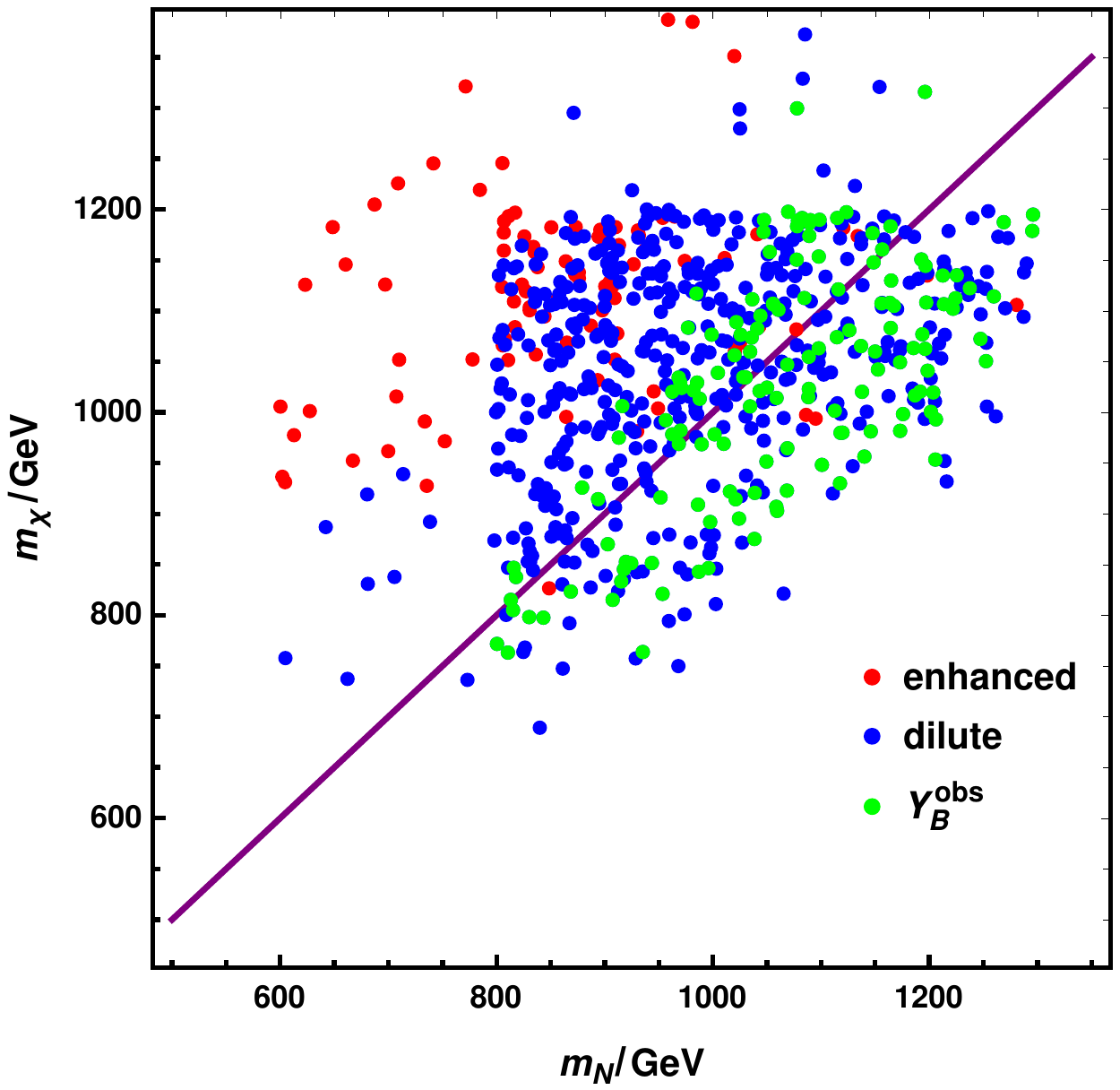}
\caption{Results of the $m_N-m_{\chi}$ satisfying dark matter relic density constraint, where the purple line represents  $m_{\chi}=m_N$. The red points represent the generated baryon asymmetry larger than the result $Y_B \approx 4.6 \times 10^{-10}$ without dark matter so that the BAU is enhanced, the blue points correspond to the case the baryon asymmetry value smaller than the reuslt without dark matter, and the green points correspond to value satisfying the observed BAU. }
 \label{Fig:14}
\end{minipage}
\centering
\begin{minipage}[t]{0.48\textwidth}
\centering
\includegraphics[width=6cm,height=6cm]{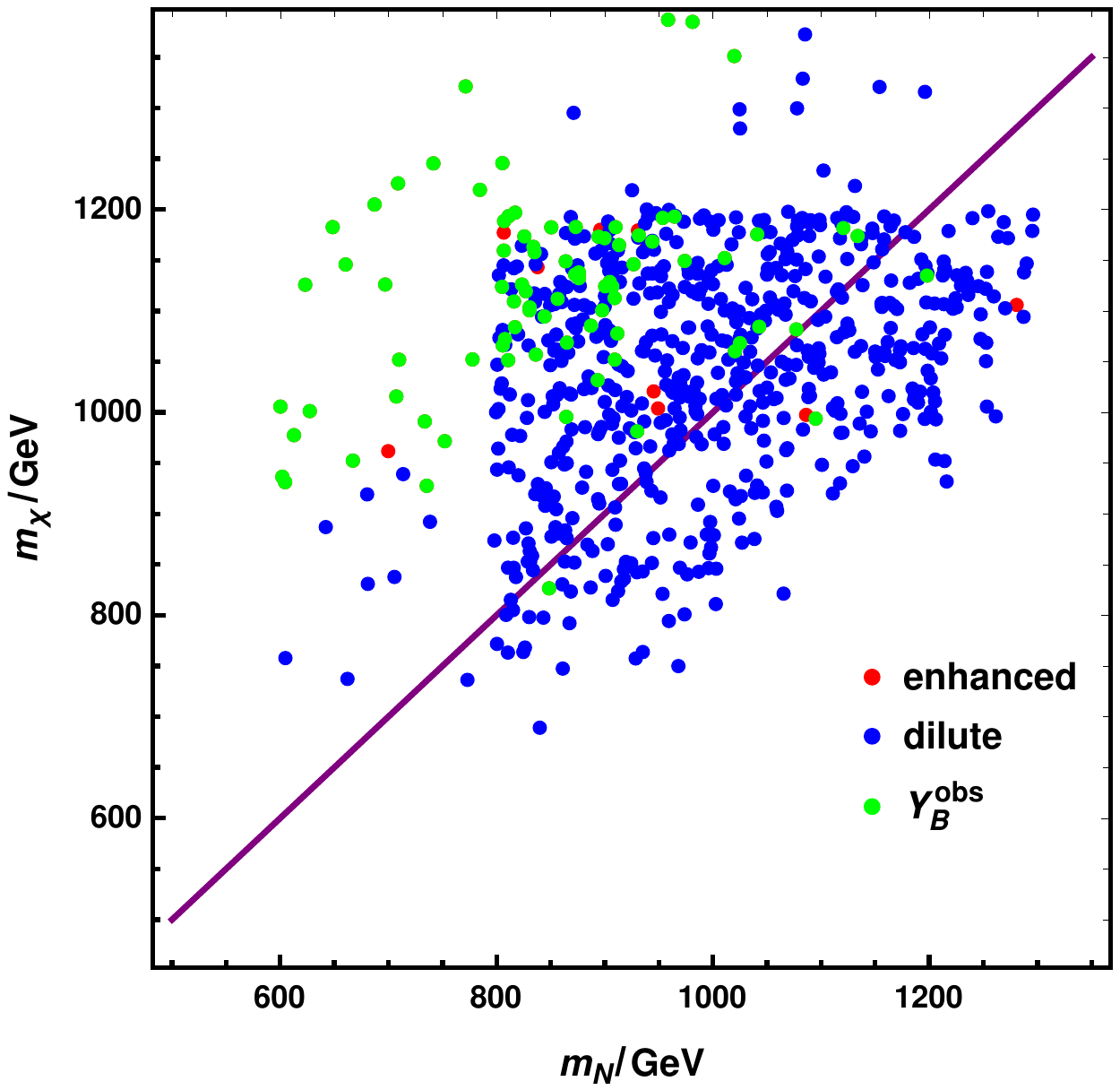}
\caption{Results of the $m_N-m_{\chi}$ satisfying dark matter relic density constraint, where the purple line represents  $m_{\chi}=m_N$. The red points represent the generated baryon asymmetry larger than the result $Y_B \approx 4.6 \times 10^{-11}$ without dark matter so that the BAU is enhanced, the blue points correspond to the case the baryon asymmetry value smaller than the reuslt without dark matter, and the green points  correspond to value satisfying the observed BAU. }
 \label{Fig:15}
\end{minipage}
\end{figure}

According to the above discussion, the interplay between dark matter and leptogenesis is more explicit in the case of $m_2=1.5m_1$ and $\sin\theta=0.9$. 
In this part, we consider the combined constraint of the  dark matter relic density and baryon asymmetry on the parameter space with $m_2=1.5m_1$ and $\sin\theta=0.9$, 
where the observed BAU is $Y_{B}^{obs}\approx 10^{-10}$ \cite{Planck:2015fie,Planck:2018vyg}. 
We consider parameters satisfying dark matter relic density in Part.~\ref{part1} and evolve the Boltzmann equations to obtain the baryon asymmetry.
On the other hand, to estimate the effect of dark matter on the BAU, we should also compare the obtained baryon asymmetry value with the result without dark matter. 
In Fig.~\ref{Fig:14} and Fig.~\ref{Fig:15}, we obtain the baryon asymmetry by fixing $\epsilon_{CP}$ different values. 
In Fig.~\ref{Fig:14}, we set $\epsilon_{CP}=10^{-5}$ and the baryon asymmetry without dark matter is about $4.6 \times 10^{-10}$  
while in Fig.~\ref{Fig:15}, we fix $\epsilon_{CP}=10^{-6}$ so that the baryon asymmetry without dark matter is about $4.6\times 10^{-11}$ within the chosen parameter space. 
Compared with the observed BAU value, it is obvious that the dilution process 
should be dominant to generate the correct BAU for $Y_{B} \approx 4.6 \times 10^{-10}$ 
and more right-handed neutrinos should be generated to obtain the right BAU in the latter case.
  
According to Fig.~\ref{Fig:14} and Fig.~\ref{Fig:15}, 
the purple line represent the relation $m_{\chi}=m_N$, 
the red (blue) points represent the generated baryon asymmetry larger (smaller) than the result without dark matter, 
and the green points are equal to the observed BAU. 
Most of the red points lie above the purple line corresponding to the region $m_{\chi}>m_N$ and the baryon asymmetry is enhanced since more right-handed neutrnios have been generated. 
On the other hand, the blue points scatter on both sides of the purple line, 
which indicates the dilution effect can make difference with suitable parameters in either $m_{\chi}>m_N$ or $m_{\chi}<m_N$ 
even though the process $NN \to \chi\chi$ is suppressed in the former case. 
Similarly, the enhanced effect can also have an impact when $m_{\chi}<m_N$ as we can see some red points lie below the purple line in Fig.~\ref{Fig:14}. 
Both situations are consistent with the discussion of Part \ref{part2}. 
In Fig.~\ref{Fig:15}, we encounter the opposite case since more right-handed neutrinos should be generated to gaurantee the correct BAU, 
and most of the green points lie in the enhanced region. 
Although the baryon asymmetry is generated via the resonant leptogenesis, 
existence of the dark matter in the model can play an important role in determing the BAU, since both dilution effect and enhanced effect can ocuur to generate the correct baryon asymmetry. 

\section{Summary}
\label{sec:6}

In this work, we discuss the interplay between dark matter and leptogenesis in a common framework. 
We do not give a UV-completion model but consider a minimal scenario including right-handed neutrino and a fermion dark matter. 
Dark matter and right-handed neutrino are connected by the mixing of two singlet scalar fields. 
We consider the decoupling limit that mixings of SM doublets with the singlet fields negligible, 
and dark matter relic density is determined by the two new Higgs particles and right-handed neutrinos. 
On the other hand, we consider the right-handed neutrino masses are  degenerate at the TeV level 
and the baryon asymmetry is generated by the resonant leptogenesis. 
As for the two singlet scalar fields, we consider two cases with $m_2=1.5m_1$ and $m_2=0.8m_1$ for simplicity. 
We scan the parameter space satisfying relic density constraint in the case of $\sin\theta=0.9$ and $\sin\theta=0.01$, 
and found a more flexible parameter space in the case of $\sin\theta=0.9$, 
where $NN \to \chi\chi$ can also play an important role in determining dark matter relic density. 
Then, we discuss the relationship between dark matter and leptogenesis in the case of $\sin\theta=0.9$ 
since dark matter will make little difference in the leptogenesis in the case of small $\sin\theta$. 
We generate the baryon asymmetry via the resonant leptogenesis, and the existence of dark matter in the model can not only dilute the baryon asymmetry result 
but may also strengthen the baryon asymmetry since more right-handed neutrnios can be generated via the  process of $\chi\chi \to NN$, and both the enhanced effect and dilution effect can occur in either case of $m_{\chi}>m_N$ and $m_{\chi}<m_N$.

\begin{acknowledgments}
\noindent
Hao Sun is supported by the National Natural Science Foundation of China (Grant No. 12075043, No. 12147205).
\end{acknowledgments}

\appendix
\section{Appendix}
   \label{App}
 We give the complete expressions of the cross section for the $\chi\chi\to h_{1,2}h_{1,2}$ in this part. For $\chi\chi \to h_1h_1$, we have 
 \begin{align}
 \sigma_{11} &=\frac{F^1_{11}+F^2_{11}+F^3_{11}+F^4_{11}}{64 \pi (-4 m_{\chi}^4 + (-2 m_{\chi}^2 + s)^2)}
 \end{align}
 where $F_{11}^i~(i=1, 2, 3, 4)$ are defined by
 \begin{align}
 F_{11}^1 &= \frac{(4m_{\chi}^2-s)\sqrt{(s-4m_1^2)(s-4m_{\chi}^2)}}{2m_{\chi}^2v_b^4}(
 -\frac{cos^4\theta sin^2\theta}{(m_2^2-s)^2}(-2 \sqrt{2} \lambda_{sx}^2 sin\theta v_b^2 (sin^2\theta-2 cos^2\theta) +  \notag \\ 
 & (m_1^2-m_2^2) 2 cos\theta \lambda_{sx} m_{\chi} v_b (2 m_1^2+m_2^2)+3 \sqrt{2} m_{\chi}^2 sin\theta (cos^2\theta m_1^2+m_2^2 sin^2\theta))^2 + \frac{6sin^2\theta cos^2\theta}{(m_1^2-s)(s-m_2^2)} \notag \\
  & (-2 \sqrt{2} \lambda_{sx}^2 sin\theta v_b^2 (sin^2\theta-2 cos^2\theta) (m_1^2-m_2^2) +2 cos\theta \lambda_{sx} m_{\chi} v_b (2 m_1^2+m_2^2)+ 3 \sqrt{2}m_{\chi}^2 \notag \\
 &  sin\theta (cos^2\theta m_1^2+m_2^2 sin^2\theta)) (-2 cos^5\theta \lambda_{sx} m_1^2 m_{\chi}v_b-2 cos^3\theta \lambda_{sx} m_1^2 m_{\chi} sin^2\theta v_b+ \notag \\
 & \sqrt{2} cos^2\theta sin^3\theta (2 \lambda_{sx}^2 v_b^2 (m_1^2-m_2^2) +m_1^2 m_{\chi}^2)+\sqrt{2}m_2^2 m_{\chi}^2 sin^5\theta)-\frac{9}{(m_1^2-s)^2} \notag \\
  & (-2 cos^5\theta \lambda_{sx} m_1^2 m_{\chi} sin\theta v_b-2 cos^3\theta \lambda_{sx} m_1^2 m_{\chi} sin^3\theta v_b +\sqrt{2} cos^2\theta sin^4\theta (2 \lambda_{sx}^2 v_b^2 (m_1^2-m_2^2)  \notag \\
 & +m_1^2 m_{\chi}^2)+\sqrt{2} m_2^2 m_{\chi}^2 sin^6\theta)^2) \\
 F_{11}^2 & = \frac{8cos^2\theta \lambda_{sx}^2 sin^3\theta}{(m_2^2-s)v_b^2}(\sqrt{2} cos^3\theta \lambda_{sx} m_{\chi} v_b (2 m_1^2+m_2^2)+ cos^2\theta sin\theta (4 \lambda_{sx}^2 v_b^2 (m_1^2-m_2^2)+3 m_1^2 m_{\chi}^2) \notag \\
 & +\sqrt{2} cos\theta \lambda_{sx} m_{\chi} sin^2 \theta v_b (2 m_1^2+m_2^2)+sin^3\theta (2 \lambda_{sx}^2 v_b^2 (m_2^2-m_1^2)+3m_2^2 m_{\chi}^2) \notag \\
 & (2 \sqrt{(s-4m_1^2) (s-4m_{\chi}^2)}-(2 m_1^2-8 m_{\chi}^2+s) \log (\frac{(\sqrt{(s-4 m_1^2) (s-4 m_{\chi}^2)}-s)+2m_1^2}{2m_1^2-(\sqrt{(s-4m_1^2) (s-4m_{\chi}^2)}+s)})) \\
 F_{11}^3 &= \frac{24\lambda_{sx}^2sin^3\theta}{(m_1^2-s)v_b^2}( \sqrt{2} cos^5 \theta \lambda_{sx} m_1^2 m_{\chi}v_b+\sqrt{2} cos^3\theta \lambda_{sx} m_1^2 m_{\chi} sin^2\theta v_b-cos^2\theta sin^3\theta (2 \lambda_{sx}^2 v_b^2 (m_1^2-m_2^2) \notag \\
&  +m_1^2 m_{\chi}^2)-m_2^2 m_{\chi}^2 sin^5\theta
)((2 m_1^2-8 m_{\chi}^2+s) \log (\frac{(\sqrt{(s-4m_1^2) (s-4 m_{\chi}^2)}-s)+2m_1^2}{2 m_1^2-(\sqrt{(s-4m_1^2)(s-4m_{\chi}^2)}+s)}) \notag \\
 & -2 \sqrt{(s-4m_1^2) (s-4m_{\chi}^2)})  \\
 F_{11}^4 & = 16 \lambda_{sx}^4 sin^4\theta (\frac{(m_1^4+4m_{\chi}^2 (s-4 m_{\chi}^2)) \log (\frac{(\sqrt{(s-4m_1^2)(s-4m_{\chi}^2)}-s)+2m_1^2}{2m_1^2-(\sqrt{(s-4m_1^2)(s-4m_{\chi}^2)}+s)})}{2 m_1^2-s}-\frac{1}{2} \sqrt{(s-4 m_1^2)(s-4 m_{\chi}^2)})
 \end{align}
 For $\chi\chi \to h_1h_2$, we have 
 \begin{align}
 \sigma_{12} &= \frac{F_{12}^1+F_{12}^2+F_{12}^3+F_{12}^4}{32 \pi ((s-2 m_{\chi}^2)^2-4m_{\chi}^4)}
 \end{align}
   where $F_{12}^i~(i=1, 2, 3, 4)$ are defined by
 \begin{align}
  F_{12}^1 & = \frac{1}{2} cos^2\theta sin^2\theta \sqrt{\frac{(s-4 m_{\chi}^2) (m_1^4-2m_1^2 (m_2^2+s)+(m_2^2-s)^2)}{s}}  \notag \\
  &(-\frac{16 \lambda_{sx}^4 (2 m_{\chi}^2 (m_1^2-m_2^2)^2+s (-8 m_{\chi}^2 (m_1^2+m_2^2)+3 m_1^2 m_2^2+16 m_{\chi}^4)+2 m_{\chi}^2 s^2)}{m_1^4 m_{\chi}^2+m_1^2 (m_2^2 (s-2 m_{\chi}^2)-2 m_{\chi}^2 s)+m_{\chi}^2 (m_2^2-s)^2}+ \notag \\ 
 & \frac{32cos\theta\lambda_{sx}^2}{(m_2^2-s)v_b^2} (3cos\theta m_{\chi}^2 (cos^2\theta m_1^2 + m_2^2 sin^2\theta) - 
  \sqrt{2}\lambda_{sx}(m_1^2 + 2 m_2^2) m_{\chi}sin\theta v_b + 2cos\theta  \notag \\ & \lambda_{sx}^2 (m_1^2-m_2^2)(cos^2\theta - 2 sin^2\theta)v_b^2)-\frac{cos^2\theta(4m_{\chi}^2}{m_{\chi}^2(m_2^2-s)^2v_b^2}(3\sqrt{2} cos\theta m_{\chi}^2 (cos^2\theta m_1^2 + m_2^2 sin^2\theta) -  \notag \\
 & 2\lambda_{sx} (m_1^2 + 2 m_2^2) m_{\chi} sin\theta v_b + 
  2 \sqrt{2} cos\theta \lambda_{sx}^2 (m_1^2 - m_2^2) (cos^2\theta - 2 sin^2\theta) v_b^2)^2 - \notag \\
 & \frac{(4m_{\chi}^2-s)sin^2\theta}{m_{\chi}^2(m_1^2-s)^2v_b^4}(3\sqrt{2} m_{\chi}^2 sin\theta (cos^2\theta m_1^2 + m_2^2 sin^2\theta) + 
  2 cos\theta \lambda_{sx} (2m_1^2 + m_2^2) m_{\chi}v_b - \notag  \\
 & 2\sqrt{2}\lambda_{sx}^2 (m_1^2 - m_2^2)sin\theta (-2 cos^2\theta + sin^2\theta) v_b^2)^2 + \frac{1}{m_{\chi}^2(m_1^2-s)(s-m_2^2)v_b^4} \notag \\
 & 2 cos\theta (4 m_{\chi}^2 - s) sin\theta (3 \sqrt{2} m_{\chi}^2 sin\theta (cos^2\theta m_1^2 + m_2^2 sin^2\theta) + 
    2 cos\theta \lambda_{sx} (2 m_1^2 + m_2^2) m_{\chi}v_b -  \notag \\
   & 2 \sqrt{2}
      \lambda_{sx}^2 (m_1^2 - m_2^2)  sin\theta (-2 cos^2\theta + 
       sin^2\theta)v_b^2) (3 \sqrt{2} cos\theta m_{\chi}^2 (cos^2\theta m_1^2 + m_2^2 sin^2\theta) -  \notag \\
   & 2 \lambda_{sx} (m_1^2 + 2m_2^2) m_{\chi}sin\theta v_b + 
    2 \sqrt{2} \lambda_{sx}^2 (m_1^2 - m_2^2)  (cos^3\theta - 
       2 cos\theta sin^2\theta) v_b^2) + \notag \\
      &  \frac{1}{(m_1^2-s)v_b^2}
 32 \lambda_{sx}^2 sin\theta (\sqrt{2} cos^3\theta \lambda_{sx} (2 m_1^2 + m_2^2) m_{\chi} v_b + \sqrt{2} cos\theta \lambda_{sx} (2m_1^2 + m_2^2) m_{\chi}sin^2\theta v_b + \notag \\
   & cos^2\theta sin\theta (3 m_1^2 m_{\chi}^2 + 4\lambda_{sx}^2 (m_1^2 - m_2^2) v_b^2) +  sin^3 \theta(3 m_2^2 m_{\chi}^2 + 2 \lambda_{sx}^2 (m_2^2-m_1^2)v_b^2)))
 \end{align}
 \begin{align}
 F_{12}^2 &= 8 cos^2\theta \lambda_{sx}^4 sin^2\theta (\frac{2 m_1^2 m_2^2-32 m_{\chi}^4+8 m_{\chi}^2 s}{m_1^2+m_2^2-s}+m_1^2+m_2^2-8 m_{\chi}^2-s) \notag \\  
 &\log (\frac{m_1^2+\sqrt{\frac{(s-4 m_{\chi}^2) (m_1^4-2 m_1^2 (m_2^2+s)+(m_2^2-s)^2)}{s}}+m_2^2-s}{m_1^2-\sqrt{\frac{(s-4 m_{\chi}^2)(m_1^4-2m_1^2 (m_2^2+s)+(m_2^2-s)^2)}{s}}+m_2^2-s})
 \end{align}
 \begin{align}
 F_{12}^3 & = -\frac{4 \sqrt{2} cos^2\theta \lambda_{sx}^2 sin^3\theta (m_1^2+m_2^2-8 m_{\chi}^2+s)}{(m_1^2-s)v_b^2}(3 \sqrt{2} m_{\chi}^2 sin\theta(cos^2\theta m_1^2 + m_2^2 sin^2\theta) + \notag \\
 & 2cos\theta \lambda_{sx} (2 m_1^2 + m_2^2) m_{\chi}v_b - 
  2 \sqrt{2}\lambda_{sx}^2 (m_1^2 - m_2^2) sin\theta (-2 cos^2\theta + sin^2\theta) v_b^2)  \notag \\
 &\log \left(\frac{m_1^2+s \left(\sqrt{\frac{\left(s-4 m_{\chi}^2\right) \left(m_1^4-2 m_1^2 \left(m_2^2+s\right)+\left(m_2^2-s\right)^2\right)}{s^3}}-1\right)+m_2^2}{m_1^2-s \left(\sqrt{\frac{\left(s-4 m_{\chi}^2\right) \left(m_1^4-2 m_1^2 \left(m_2^2+s\right)+\left(m_2^2-s\right)^2\right)}{s^3}}+1\right)+m_2^2}\right)
 \end{align}
 \begin{align}
 F_{12}^4 & = \frac{4 \sqrt{2} cos^3\theta \lambda_{sx}^2 sin^2\theta \left(m_1^2+m_2^2-8 m_{\chi}^2+s\right)}{v_b^2 \left(m_2^2-s\right)}(-2 \sqrt{2} \lambda_{sx}^2 v_b^2 \left(cos^3\theta-2 cos\theta sin^2\theta\right) (m_1^2-m_2^2)+  \notag \\
 & 2\lambda_{sx}m_{\chi}sin\theta v_b  \left(m_1^2+2 m_2^2\right)-3 \sqrt{2} cos\theta m_{\chi}^2 \left(cos^2\theta m_1^2+m_2^2 sin^2\theta\right) )\notag \\
 & \log \left(\frac{m_1^2+s \left(\sqrt{\frac{\left(s-4 m_{\chi}^2\right) \left(m_1^4-2 m_1^2 \left(m_2^2+s\right)+\left(m_2^2-s\right)^2\right)}{s^3}}-1\right)+m_2^2}{m_1^2-s \left(\sqrt{\frac{\left(s-4 m_{\chi}^2\right) \left(m_1^4-2 m_1^2 \left(m_2^2+s\right)+\left(m_2^2-s\right)^2\right)}{s^3}}+1\right)+m_2^2}\right)
 \end{align}
  For $\chi\chi \to h_2h_2$, we have 
 \begin{align}
   \sigma_{22} &= \frac{F_{22}^1+F_{22}^2+F_{22}^3+F_{22}^4}{64 \pi (-4m_{\chi}^4 + (-2 m_{\chi}^2 + s)^2)}
 \end{align}
   where $F_{22}^1, F_{22}^2, F_{22}^3, F_{22}^4$ are defined by
 \begin{align}
   F_{22}^1 &= \frac{(4m_{\chi}^2 - s)  \sqrt{(-4 m_2^2 + s) (-4 m_{\chi}^2 + s)} }{2m_{\chi}^2v_b^4}(
  -\frac{sin^4\theta }{(m_1^2-s)^2}(3 \sqrt{2} cos^2\theta m_{\chi}^2 (cos^2\theta m_1^2+m_2^2 sin^2\theta)+  \notag \\
  & 2 \sqrt{2} \lambda_{sx}^2 v_b^2 (cos^4\theta-2 cos^2\theta sin^2\theta) (m_1^2-m_2^2)-2 cos\theta \lambda_{sx} m_{\chi} sin\theta v_b (m_1^2+2 m_2^2))^2 + \notag \\       & \frac{6cos^2\theta sin^2\theta}{(m_1^2-s)(s-m_2^2)}(2 \sqrt{2} \lambda_{sx}^2 v_b^2 (cos^3\theta-2 cos\theta sin^2\theta) (m_1^2-m_2^2)-2 \lambda_{sx} m_{\chi}sin\theta v_b (m_1^2+2 m_2^2)+ \notag \\
 & 3 \sqrt{2} cos\theta m_{\chi}^2 (cos^2\theta m_1^2+m_2^2 sin^2\theta)) (\sqrt{2} cos^5\theta m_1^2 m_{\chi}^2+\sqrt{2} cos^3\theta sin^2\theta (2 \lambda_{sx}^2 v_b^2 (m_2^2-m_1^2)+m_2^2 m_{\chi}^2)  \notag \\
 & +2 cos^2\theta \lambda_{sx} m_2^2 m_{\chi} sin^3\theta v_b+2 \lambda_{sx} m_2^2 m_{\chi}sin^5\theta v_b)-\frac{9}{(m_2^2-s)^2}(\sqrt{2} cos^6\theta m_1^2 m_{\chi}^2 + \notag \\
 & 2 cos^3\theta \lambda_{sx} m_2^2 m_{\chi} sin^3\theta v_b +  
  2 cos\theta \lambda_{sx} m_2^2 m_{\chi} sin^5\theta v_b + 
  \sqrt{2} cos^4\theta sin^2\theta (m_2^2 m_{\chi}^2 + 
     2 \lambda_{sx}^2 (m_2^2-m_1^2) v_b^2))^2)
 \end{align}
 \begin{align}
   F_{22}^2 &=\frac{24cos^3\theta\lambda_{sx}^2}{(m_2^2-s)v_b^2}(cos^3\theta m_{\chi}^2 (cos^2\theta m_1^2 + m_2^2 sin^2\theta) + 
  \sqrt{2} \lambda_{sx} m_2^2 m_{\chi} sin^3\theta v_b + 
  2 cos^3\theta \lambda_{sx}^2 (-m_1^2 + m_2^2) sin^2\theta v_b^2) \notag \\
  & (2 \sqrt{(-4 m_2^2 + s) (-4 m_{\chi}^2 + s)} - (2 m_1^2 - 8 m_{\chi}^2 + s) \log(\frac{
    2 m_1^2 + (-s + \sqrt{(-4 m_2^2 + s) (-4 m_{\chi}^2 + s)})}{
    2 m_1^2 - (s + \sqrt{(-4 m_2^2 + s) (-4 m_{\chi}^2 + s)})})))
 \end{align}
 \begin{align}
  F_{22}^3 &= 8 cos^4\theta \lambda_{sx}^4 ((2 m_2^2-8 m_{\chi}^2-s) \log (\frac{(\sqrt{(s-4 m_2^2) (s-4 m_{\chi}^2)}-s)+2 m_2^2}{2 m_2^2-(\sqrt{(s-4 m_2^2) (s-4m_{\chi}^2)}+s)})- \notag \\
  & \frac{\sqrt{(s-4 m_2^2) (s-4 m_{\chi}^2)} (2 (m_2^4-6 m_2^2 m_{\chi}^2+8 m_{\chi}^4)+m_{\chi}^2 s)}{m_2^4-4 m_2^2 m_{\chi}^2+m_{\chi}^2 s})
 \end{align}
 \begin{align}
  F_{22}^4 &= 8cos^4\theta \lambda_{sx}^4 (-\sqrt{(s-4 m_2^2) (s-4m_{\chi}^2)}+\frac{1}{m_1^2+m_2^2-s}(m_1^4-m_1^2 s+m_2^2 s-16 m_{\chi}^4+4 m_{\chi}^2 s) \notag \\
 & \log (\frac{2 m_1^2+(\sqrt{(s-4 m_2^2) (s-4 m_{\chi}^2)}-s)}{2m_1^2-(\sqrt{(s-4 m_2^2)(s-4m_{\chi}^2)}+s)})+(m_2^4+4 m_{\chi}^2 (s-4 m_{\chi}^2)) \notag \\
 & \log (\frac{(\sqrt{(s-4m_2^2)(s-4m_{\chi}^2)}-s)+2 m_2^2}{2 m_2^2-(\sqrt{(s-4m_2^2) (s-4m_{\chi}^2)}+s)}))
 \end{align}
\bibliographystyle{JHEP}
\bibliography{v1.bib}

\end{document}